\pdfoutput=1

\documentclass[letter, twocolumn]{aastex631}

\usepackage{epstopdf}
\usepackage{xspace}
\usepackage{graphicx}
\usepackage{amssymb}
\usepackage{amsmath}
\usepackage{natbib}
\usepackage{float}
\usepackage{hyperref}
\newcommand\chandra{{\it Chandra}}
\newcommand\fermi{{\it Fermi}}

\begin{document}

\title{On the Origin of the X-ray Emission in Heavily Obscured Compact Radio Sources}

\author[0000-0002-3626-5831]{Dominika ~{\L}.~Kr\'{o}l}
\affiliation{Astronomical Observatory of the Jagiellonian University, Orla 171, 30-244 Krak\'{o}w, Poland}
\affiliation{Center for Astrophysics $|$  Harvard \& Smithsonian, 60 Garden Street, Cambridge, MA 02138, USA}
\author[0000-0002-6286-0159]{Ma{\l}gosia Sobolewska}
\affiliation{Center for Astrophysics $|$  Harvard \& Smithsonian, 60 Garden Street, Cambridge, MA 02138, USA}

\author[0000-0001-8294-9479]{{\L}ukasz Stawarz}
\affiliation{Astronomical Observatory of the Jagiellonian University, Orla 171, 30-244 Krak\'{o}w, Poland}

\author[0000-0002-0905-7375]{Aneta Siemiginowska}
\affiliation{Center for Astrophysics $|$  Harvard \& Smithsonian, 60 Garden Street, Cambridge, MA 02138, USA}

\author[0000-0003-0216-8053]{Giulia Migliori}
\affiliation{Istituto di Radioastronomia – INAF, Via P. Gobetti 101, I-40129 Bologna, Italy}
\affiliation{Dipartimento di Fisica e Astronomia, Universit `a di Bologna, Via Gobetti 93/2, I-40129 Bologna, Italy}

\author[0000-0003-0406-7387]{Giacomo Principe}
\affiliation{Istituto di Radioastronomia – INAF, Via P. Gobetti 101, I-40129 Bologna, Italy}
\affiliation{Universitá di Trieste, Dipartimento di Fisica, I-34127 Trieste, Italy}
\affiliation{Istituto Nazionale di Fisica Nucleare, Sezione di Trieste, I-34127 Trieste, Italy}

\author[0000-0003-0685-3621]{Mark A. Gurwell}
\affiliation{Center for Astrophysics $|$  Harvard \& Smithsonian, 60 Garden Street, Cambridge, MA 02138, USA}

\begin{abstract}

X-ray continuum emission of active galactic nuclei (AGNs) may be reflected by circumnuclear dusty tori, producing prominent fluorescence iron lines at X-ray frequencies. Here we discuss the broad-band emission of three radio-loud AGN belonging to the class of compact symmetric objects (CSOs), with detected narrow Fe\,K$\alpha$ lines. CSOs have newly-born radio jets, forming compact radio lobes with projected linear sizes of the order of a few to hundreds of parsecs. We model the radio--to--$\gamma$-ray spectra of compact lobes in {J1407+2827}, J1511+0518, and {J2022+6137},  which are among the nearest and the youngest CSOs known to date, and are characterized by an intrinsic X-ray absorbing column density of  $N_{\rm H} > 10^{23}$\,cm$^{-2}$. In addition to the archival data, we analyze the newly acquired \chandra\ X-ray Observatory and Sub-Millimeter Array (SMA) observations, and also refine the $\gamma$-ray upper limits from the \fermi\ Large Area Telescope (LAT) monitoring. The new \chandra\ data exclude the presence of the extended X-ray emission components on scales larger than $1.5^{\prime \prime}$. The SMA data unveil a correlation of the spectral index of the electron distribution in the lobes and $N_{\rm H}$, which can explain the $\gamma$-ray quietness of heavily obscured CSOs. Based on our modeling, we argue that the inverse-Compton emission of compact radio lobes may account for the intrinsic X-ray continuum in all these sources. Furthermore, we propose that the observed iron lines may be produced by a reflection of the lobes' continuum from the surrounding cold dust.
\end{abstract}

\section{Introduction} \label{sec:intro}
Hard X-ray emission of active galactic nuclei (AGNs) is typically associated with the presence of a corona, a hot plasma in the accretion disk vicinity, where optical/UV disk photons gain energy via the process of Comptonization  \citep{Haardt91,Haardt93,Svensson94,Stern95,Esin97}. Throughout the years, a variety of models have been discussed to describe coronas' location and structure, including a lamp post model \citep{Esin97,Yuan04}, a hot inner flow \citep{Done99, Yuan14}, or a jet/base of a jet \citep{Markoff05,Dauser13}. The X-ray continuum is often accompanied by fluorescent iron lines, which are now established as ubiquitous features in the X-ray spectra of both obscured and unobscured AGNs \citep[e.g.,][]{Page04,Nandra07,Ricci14}.
A narrow Fe\,K$\alpha$ line is produced when X-ray continuum emission reflects off mater located either on the line-of-sight (e.g., in the broad line region clouds), or out of the line-of-sight, e.g., dusty torus or outer regions of an accretion disk. In the latter case, the iron line is accompanied by a Compton reflection component. \citep[e.g.,][]{George91,Matt96,Nandra06,Bianchi08,Shu10,Gandhi15}.

In this paper, we complement the above scenarios by proposing young, compact lobes of radio galaxies and quasars as another source of AGN X-ray continuum emission. These lobes constitute a reservoir of a hot magnetized plasma back-flowing from the jet termination shock, where the jet's bulk kinetic energy is converted to the internal energy of the jet particles \citep[e.g.,][]{Begelman89}. Unlike in the case of relativistic beamed jets, the lobes' emission is isotropic in the observers' rest frame. During the earliest phases of the jet lifetime, the lobes have a size that is smaller or comparable to that of the dusty torus.  Consequently, in objects where the primary X-ray continuum originates from radio lobes expanding inside or near a molecular torus, one should expect to observe X-ray reflection and fluorescence features, such as the Fe K$\alpha$ line.

With this picture in mind, we model the broad-band spectral energy distributions (SEDs) of three objects: {{J1407+2827} (also known as OQ+208 or Mrk 668), J1511+0518,  and {J2022+6137 (2021+614)}}, to quantitatively explore the validity of the above expectation. These objects
were selected from a sample of compact symmetric objects \citep[CSOs, a subset of AGNs known for their compact, young radio structures, see e.g.,][]{ODea98,ODea21,Kiehlmann23a,Kiehlmann23b,Readhead23} considered in \cite{Sobolewska19a} based on  their distinctive X-ray spectral features derived based on  extensive observations by \chandra, XMM-{\it Newton}, and {\it NuSTAR} \citep[see][further denoted as S19 and S23 respectively]{Sobolewska19b,Sobolewska23}. Key features are a high intrinsic X-ray absorption (equivalent hydrogen column densities, $N_{\rm H}$\,$>\,10^{23}$\,cm$^{-2}$), as well as a narrow fluorescent Fe\,K$\alpha$ line and the broad Compton reflection component indicating a reflection from a toroidal obscurer. Such properties make the selected targets rare examples of ``radio-loud/X-ray--obscured'' AGNs \citep[see][]{Panessa16,LaMassa23}.

The observed radio morphology of these sources, at the milliarcsecond scale, comprises a weak compact core and symmetric lobes, with the innermost projected linear sizes within the 7--25\,pc range \citep{An12}. These sizes are comparable to the radii of dusty molecular tori in nearby Seyfert galaxies, as measured by the Atacama Large Millimeter Array (ALMA)  \citep[][]{Combes19,Garcia21}. The integrated radio spectra display spectral turnovers at GHz frequencies, suggesting excess absorption in the lobes' radio continuum emission \citep{Wojtowicz20,Kiehlmann23a}. %These radio characteristics classify them as Compact Symmetric Objects (CSOs), a subset of AGNs known for their compact, young radio structures \citep[see][]{ODea98,ODea21,Kiehlmann23a,Kiehlmann23b,Readhead23}. 
Additionally, the mid-infrared colors of these sources, as measured by the Wide-field Infrared Survey Explorer (WISE), place them in the ``AGN'' classification meaning that their observed radiative output at the $\mu$m range is indeed dominated by the circumnuclear dust, typical for quasars and luminous Seyfert galaxies \citep[see][]{Kosmaczewski20,Nascimento22}.

To investigate the potential of radio lobes as a source of X-ray continuum in the above mentioned CSOs, we constructed and then modeled the broadband SED using archival radio, infrared (IR), optical/UV, and X-ray data. Additionally, we incorporated recent sub-millimeter data collected with the SMA. Finally, we revisit the \fermi-LAT data, which indicate that our three CSOs are $\gamma$-ray--quiet \citep{Principe20}, and derive $\gamma$-ray upper limits based on 14 years of LAT monitoring. Our modeling approach employs a simple dynamical description of the expanding lobes in young radio  sources. The lobes' X-ray emission is produced via inverse-Compton (IC) scattering of various soft photon fields -- predominantly those emitted by accretion disks and dusty tori -- by ultra-relativistic electrons injected by the jets into evolving compact lobes. All relevant energy losses of the radiating elctrons are taken into account and integrated over the entire lifetime of the source \citep[][further referenced as LS08]{Stawarz08}. We show that in the case of all three sources X-ray emission can be successfully modeled within this scenario.

In addition, we present the new X-ray data collected with \chandra\ for J1511+0518. These new data enable us to perform an image analysis of the source and investigate the presence of an extended X-ray emission on scales larger than $1.5^{\prime \prime}$($\sim$\,$2.4$\,kpc). Let us note that, in contrast to our heavily X-ray absorbed $\gamma$-ray quiet CSOs, the three $\gamma$-ray loud CSOs detected to date with \fermi-LAT are unobscured in X-rays. These include PKS\,1718--649 \citep{Migliori16,Siemiginowska16}, TXS\,0128+554 \citep{Lister20}, and NGC\,3894 \citep{Principe20,Balasubramaniam21}, with the observed (isotropic) $\gamma$-ray luminosities $L_{\rm 0.1-1,000\,GeV} \simeq 1 \times 10^{42}$\,erg\,s$^{-1}$, $2 \times 10^{43}$\,erg\,s$^{-1}$, and $6 \times 10^{41}$\,erg\,s$^{-1}$, respectively \citep{Principe21}. We also note that the X-ray emission extended beyond the innermost $1.5^{\prime \prime}$ nuclear region from the radio source position has been detected around two of three $\gamma$-ray--loud CSOs \citep[PKS\,1718--649 and NGC\,3894, see details in][respectively]{Siemiginowska16,Beuchert18,Balasubramaniam21}. In both sources, this emission has been modeled as originating from a collisionally ionized plasma with temperatures of  $\sim (0.7 - 0.8)$\,keV, possibly with an additional photoionized plasma component in PKS\,1718--649 \citep{Beuchert18}. 

To date, no extended X-ray emission has been spatially resolved around the other two heavily obscured CSOs, {as discussed in S19 for {J1407+2827} and in \citet[][]{Siemiginowska16} for {J2022+6137}.} The low number of counts in the archival $2$\,ks \chandra\ observation of J1511+0518 did not allow for a meaningful spatial analysis. However, S23 pointed out that including a thermal plasma component with a temperature $\sim 1$\,keV could alleviate the residuals in the soft range of the XMM-{\it Newton} and {\it NuSTAR} joint spectrum. Since the spatial resolution of the XMM-Newton data was insufficient to determine whether this thermal emission originated within the unresolved AGN region or on scales exceeding several kpc, we reobserved J1511+0518 with \chandra\ to perform an image analysis of this source and investigate the presence of the extended emission. Here we show, that in the new $55$\,ks \chandra\ observation we do not detect any extended emission. 

This paper is organized as follows. In Section~\ref{sec:chandra}, we analyze the new \chandra\ data for J1511+0518, and show that the X-ray emission of this source is unresolved on arcsec scales. In Section~\ref{sec:multi}, we describe the multiwavelength data used to build the broadband SEDs of {{J1407+2827}, J1511+0518, and {J2022+6137}}. In Section~\ref{sec:model} we briefly summarize the applied radiative model for expanding radio lobes, and delineate the modeling procedure. We present our modeling results in Section~\ref{sec:results}. The main findings derived from the analysis are discussed further in Section~\ref{sec:discussion}, and summarized in Section~\ref{sec:summary}.

Throughout the paper, we use modern flat $\Lambda$CDM cosmology with $H_0 = 69.3$\,km\,s$^{-1}$\,Mpc$^{-1}$ and $\Omega_m = 0.287$ \citep{Hinshaw13}. {For {J1407+2827}, $z=0.076$ \citep{Stanghellini93} gives $D_{\rm L} \simeq 348$\,Mpc and $1.5$\,kpc\,arcsec$^{-1}$. For J1511+0518, its redshift $z = 0.084$ \citep{Ahn12} translates to the luminosity distance $D_{\rm L} \simeq 388$\,Mpc and the scale of $1.6$\,kpc\,arcsec$^{-1}$. Finally, for {J2022+6137}, $z=0.227$ \citep{Polatidis03} translates into $D_{\rm L} \simeq 1,144$\,Mpc and $3.7$\,kpc\,arcsec$^{-1}$}.

\section{Chandra observations of J1511+0518}
\label{sec:chandra}

We analyzed the recent (May 2022) observations of J1511+0518 with the \chandra's Advanced CCD Imaging Spectrometer (ACIS), consisting of three separate pointings summarized in Table\,\ref{tab:tab1}. 

We used the standard procedures to reduce the data with {\tt CIAO v4.14} \citep{Fruscione06}, starting with the reprocessing performed with the {\tt chandra\_repro} script. The astrometry correction was carried out with respect to the longest observation, ObsID 25457, with the {\tt wavdetect}, {\tt wcs\_match}, and {\tt wcs\_update} scripts. We modeled the \chandra\ spectra of J1511+0518 using {\tt Sherpa} \citep{Freeman01}, and the modeling details are described in Section~\ref{sec:spectral}. Next, we used the best fitting spectral model as an input to the point spread function (PSF) simulations of an on-axis point-like source using the {\tt ChaRT} \citep{Carter03} and {\tt MARX} \citep{Davis12} tools. We compared the simulated PSF with that built based on our observations to asses the presence of an extended X-ray emission on scales larger than $\sim 1.5^{\prime \prime}$ from the source coordinates (see Section~\ref{sec:psf}).

\subsection{Spectral Modeling}
\label{sec:spectral}

The source and background energy spectra were extracted separately for each observation of J1511+0518. We chose a circular source extraction region with the radius of 5 ACIS-S pixels\footnote{For ACIS-S, $1\,{\rm px} = 0.492^{\prime \prime}$.}, corresponding to $\sim 2.5^{\prime \prime}$, centered at RA = 15h\,11m\,41.18s, Dec = +5$^{\circ}$\,18$^{\prime}$\,10$^{\prime \prime}$.150 (J2000). For the background region, we chose a concentric ring with $r_{\rm in} = 10\,{\rm px} \sim 5^{\prime \prime}$\ and $r_{\rm out}=20\,{\rm px} \sim 10^{\prime \prime}$, centered at the same coordinates as the source region. The number of net source counts for each observation can be found in Table\,\ref{tab:tab1}.  We binned the spectra requiring the signal-to-noise ratio (SNR) of 3 in each bin. The maximum bin length was set to 0.2\,keV. We performed a joint spectral fit of all three \chandra\ observations in the $0.5-7.0$\,keV energy band in order to inform the PSF simulations performed in Section~\ref{sec:psf}, as part of the image analysis. 

\begin{deluxetable}{cccccc}[t]
\label{tab:tab1}
\tablecaption{\chandra\ observations of J1511+0518.}
\tablehead{\colhead{ObsID} & \colhead{Date} & \colhead{Exposure$^a$} & \multicolumn{2}{c}{Net counts in energy range$^b$} \\ \colhead{} & \colhead{} & \colhead{s} & \colhead{$0.5-7$\,keV} & \colhead{$0.5-1$\,keV}}
\startdata
    25457 & 12-May-2022 & 20,170 & 252 $\pm$ 16 &  12 $\pm$ 3 \\
    25863 & 17-May-2022 & 15,700 & 172 $\pm$ 13 & 11 $\pm$ 3 \\
    26417 & 18-May-2022 & 19,700 & 224 $\pm$ 15 & 11 $\pm$ 3 \\
\hline
\enddata
\tablecomments{$^a$ACIS-S3 exposure in seconds. $^b$Net counts within the $5\,{\rm px} \simeq 2.5^{\prime \prime}$ circular source extraction region centered at the source coordinates.}
\end{deluxetable} 
%\colhead{Net counts$^b$} &\colhead{}

\begin{figure*}
    \centering
    \includegraphics[width=0.43\textwidth]{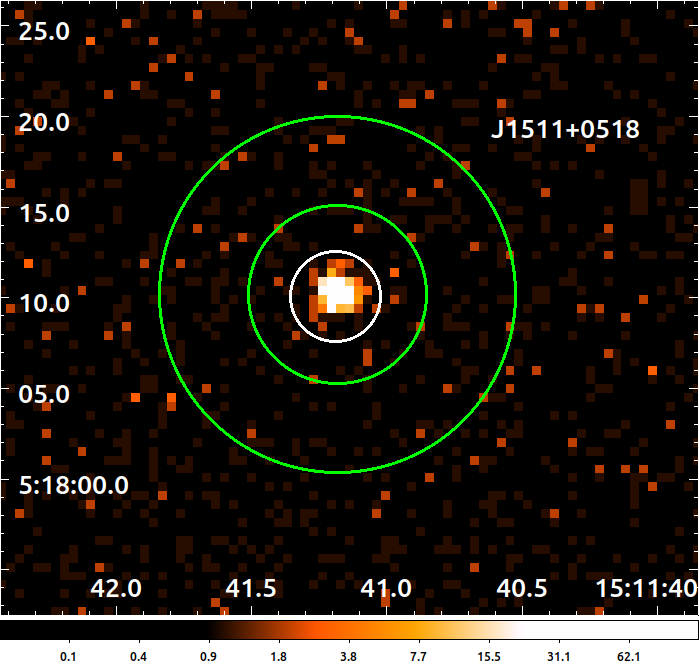} 
    \includegraphics[width=0.56\textwidth]{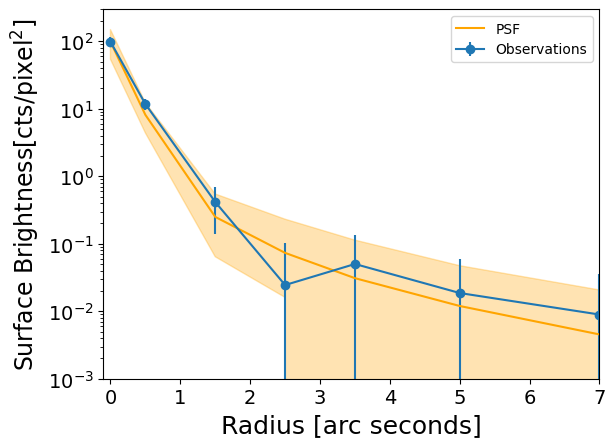} 
    \caption{Left: \chandra\ The merged {Chandra} ACIS-S image of three J1511+0518 observations. The source and background extraction regions are denoted by a circle with the radius of $r \simeq 2.5^{\prime \prime}$ (white), and an annulus with $r\simeq (5-10)^{\prime \prime}$ (green), respectively. Right: Brightness profile of the merged image (blue data points and lines), along with the simulated averaged PSF (orange curve). The orange shaded region around the average PSF indicate the range (i.e., the minimum and maximum values) observed at a given radius in $150$ random PSF realizations.}
    \label{fig:merged}
\end{figure*}

Our simplified spectral model applied  to the new \chandra\ data was based on the joint modeling of the XMM-{\it Newton} and {\it NuSTAR} data of the source by S23. In particular, our continuum model consisted of a combination of a collisionally ionized plasma, {\tt xsapec}, and a non-thermal emission approximated with a simple {\tt powerlaw} model. We fixed the photon index and the plasma temperature at $\Gamma = 1.8$ and $kT = 1.0$\,keV to avoid degeneracy in the fit parameters. Additionally, our model included a Gaussian line, {\tt xszgauss}, to account for the narrow Fe\,K$\alpha$ feature reported in S23, which is also present in the new \chandra\ data. The rest-frame energy and the width of the line were fixed at 6.4\,keV and 0.02\,keV respectively. Moreover, we have included the Galactic absorption, {\tt xsphabs}, with the equivalent hydrogen column density fixed at $N_{\rm H,\,Gal} = 3.29 \times 10^{20}$\,cm$^{-2}$ \citep{Dickey90} acting on all the spectral components, and an intrinsic absorption at the redshift of the source, {\tt xszphabs}, with the column density $N_{{\rm H},\,z}$, acting exclusively on the power-law component.

The final model, {\tt xsphabs} $\times$ {\tt (xszphabs} $\times$ {\tt powlaw1d + xsapec + xszgauss)}, had four free model parameters, which we linked across the three data sets (corresponding to the three separate pointings) during the fitting procedure. We obtained a satisfactory fit with the {\tt chi2gehrels}\footnote{\url{https://cxc.cfa.harvard.edu/sherpa/ahelp/chi2gehrels.html}} fit statistic of $43$ for $97$ degrees of freedom for the source intrinsic absorbing column density $N_{{\rm H},\,z} = (1.28 \pm 0.31) \times 10^{22}$\,cm$^{-2}$, the normalization of the power-law component, $N_{\rm PL} = (4.47 \pm 0.048) \times 10^{-5}$\,ph\,keV$^{-1}$\,cm$^{-2}$\,s$^{-1}$, the normalization of the iron line $N_{\rm Fe} = (1.43 \pm 0.69) \times 10^{-6}$\,ph\,cm$^{-2}$\,s$^{-1}$, and the normalization of the thermal plasma $N_{\rm APEC} = (4.47 \pm 0.47) \times 10^{-19} \, (1+z)^2 \, \int \textrm{d}V \, n_e n_{\rm H} / 4\pi D_{\rm L}^2$, where $\textrm{d}V$ is the emission volume element, and $n_e$ and $n_{\rm H}$ are the electron and hydrogen densities, respectively, all in cgs units. All the uncertainties represent $1 \sigma$ confidence intervals.

The best fit model returned the $0.5 - 7.0$\,keV de-absorbed flux of the power-law component $F_{\rm 0.5-7.0\,keV} \simeq (2.14 \pm 0.14) \times 10^{-13}$\,erg\,cm$^{-2}$\,s$^{-1}$, and the extrapolated 2--10\,keV power-law luminosity of $L_{\rm 2-10\,keV} \sim 2.5 \times 10^{42}$\,erg\,s$^{-1}$; the equivalent width of the iron line read as EW\,$\simeq 0.72_{-0.36}^{+1.12}$\,keV. These values are all comparable within the errors to the corresponding parameter values obtained by S23. However, we note a difference in the complexity of the models applied therein and in our work due to a much limited energy range and photon statistics of the new \chandra\ data when compared with the joint XMM-{\it Newton} and {\it NuSTAR} data sets.

\subsection{Image Analysis}
\label{sec:psf}

We simulated the \chandra\ PSF for all three pointings using the {\tt chart} tool and assuming the best-fit spectral model as presented in Section~\ref{sec:spectral} above. We repeated the simulations 50 times for each of the three observations to account for the possibility of considerable differences in each realization of the PSF due to random photon fluctuations. We used the {\tt marx} software to convert the obtained rays to the pseudo-event file with default ASPECT blur 0.07. Finally, we averaged the simulation results for each observation\textrm{s}. 

We extracted all the detected counts from concentric rings defined by the difference between the outer and inner radii of $\Delta r_{int} = (0.5 + n)^{\prime \prime}$, with $n = 0,..,4$, and $\Delta r_{ext} = (0.5 + 2n)$ with $n = 2,...,7$. A comparison of the resulting source brightness profile for the merged three pointings and the simulated averaged PSF is presented in Figure\,\ref{fig:merged}. As shown, the observed brightness profile is consistent with the emission of a point-like source within the ACIS spatial resolution, given a relatively large spread observed in various PSF realizations due to random photon fluctuations, especially at larger radii from the source coordinates (see the orange shaded area in the figure) and the additional uncertainty in the astrometry correction between the three analyzed \chandra\ pointings.

The lack of the extended emission component in the new \chandra\  data can be significantly affected by the \chandra's ACIS-S chip degradation. As the extended emission component is expected to originate from thermal hot plasma radiation, it would predominantly appear in the \chandra\ soft band, which is most affected by the chip degradation.  In this context, it is worth noting that the \chandra\ observations of CSO NGC\,3894, which revealed source' extended emission, were conducted during Cycle 9. The effective area of the ACIS-S chip has significantly decreased between Cycle 9 and 23, the latter being when J1511+618 was observed. Namely, in soft X-ray band has decreased by $\sim2$ orders of magnitude\footnote{for the comparison see \url{https://cxc.cfa.harvard.edu/cgi-bin/prop_viewer/build_viewer.cgi?ea}}.

\section{Multiwavelength Data}
\label{sec:multi}

\subsection{Sub-Millimeter Array Observations}
\label{subsec:SMA}

The Submillimeter Array was utilized to observe young radio sources and flat spectrum radio quasars (FSRQs) under the SMA programs 2018A-S042 and 2018B-S037. These programs were executed multiple times to accommodate a wide range of target locations, and thus {J1407+2827 was observed twice (8 August 2018 and 2 May 2019), J1511+0518 was observed twice (13 July and 8 August, 2018) and J2022+6137} was observed once (13 July 2018). For all the observations, complex-valued measurements (``visibilities'') of the sources were bracketed by measurements of nearby strong radio-loud AGNs (often blazars) as amplitude and phase gain calibrators. Data reduction was performed with the SMA in-house reduction package MIR\footnote{\url{https://lweb.cfa.harvard.edu/~cqi/mircook.html}}, including data flagging, initial system temperature calibration, removal of correlator-based amplitude spikes at specific channels, and generation of a broad-band \emph{continuum} channel (8\,GHz wide, in each of two sidebands). Complex gain corrections, determined from the calibrator, were applied to the target visibilities, and the flux density scale was set using a solar system standard (Callisto) and a secondary standard, MWC349a.  The calibrated continuum-band visibility data were then vector averaged to determine the flux density of each source. The SMA observations and the resulting fluxes are listed in Appendix\,\ref{sec:appendix_data}. 

\subsection{Fermi-LAT Monitoring}
\label{subsec:fermi}
The Large Area Telescope (LAT) is the main instrument on board the \fermi\ \textit{Gamma-ray Space telescope}. It is a $\gamma$-ray telescope sensitive at energies ranging from 20 MeV to more than 300 GeV \citep{Atwood09}.

In this work, we performed a dedicated analysis of the 14 years of LAT observations (between 4 August 2008 and 4 August 2022) of the three CSOs, following the analysis technique of \citet{Principe21}. We selected P8R3 SOURCE class events \citep{Bruel18}, in the energy range between 100\,MeV and 1\,TeV, from regions of interest (ROIs) of 15$^{\circ} \times 15^{\circ}$ centered at the positions of each selected source. The value of the low energy threshold is motivated by the large uncertainties in the arrival directions of the photons below 100\,MeV \citep{Principe18}, leading to a possible confusion between point-like sources and the Galactic diffuse component.

The analysis consisting of model optimization, localization, spectrum and variability study, was performed with {\tt Fermipy}\footnote{version 1.0.1 \url{http://fermipy.readthedocs.io/en/latest/}} \citep{Wood17}, a Python package that facilitates analysis of the LAT data with the \fermi\ Science Tools.

We created the count maps with a pixel size of $0.1^{\circ}$, and we excluded all $\gamma$-rays with zenith angle larger than $95^{\circ}$ in order to limit the contamination from secondary $\gamma$-rays from the Earth’s limb \citep{Abdo09}. Following the analysis technique reported in \citet{Principe20,Principe21}, we made an even harder cut at low energies ($<300$\,MeV) by reducing the maximum zenith angle ($<85^{\circ}$) and by excluding event types with the largest point spread function (PSF0).\footnote{A measure of the quality of the direction reconstruction is used to assign events to four quartiles. $\gamma$-rays in Pass 8 data can be separated into 4 PSF event types: 0, 1, 2, 3, where PSF0 has the largest point spread function and PSF3 has the best.}

We used the {\tt P8R3\_SOURCE\_V3} instrument response functions (IRFs). The selected spectral model included all the point-like and extended LAT sources located at a distance $<20^{\circ}$ from the investigated CSOs, as listed in the Fourth \fermi-LAT Source Catalog \citep[4FGL-DR2;][]{Abdollahi20}, as well as the Galactic diffuse and the isotropic emission models\footnote{\url{https://fermi.gsfc.nasa.gov/ssc/data/access/lat/BackgroundModels.html}, in particular the diffuse model {\tt gll\_iem\_v07.fits} and the isotropic component {\tt iso\_P8R3\_SOURCE\_V3\_v1.txt} .} adopted to compile the 4FGL-DR2. We first optimized the model for a given ROI, then we searched for additional new sources not included in 4FGL-DR2, and finally, if the source was significantly detected ($>3 \sigma$ significance), we re-localized the source.

During the model fitting, we left the normalization of the isotropic and Galactic diffuse backgrounds as well as the spectral parameters of the sources within 5$^{\circ}$ of our targets free to vary, while for the sources at a distance between 5$^{\circ}$ and 10$^{\circ}$, only the normalization was fitted; we fixed the parameters of all the sources within the ROIs at larger angular distances from our targets to their 4FGL values. Spectral fits were performed over the energy range from 100\,MeV to 1\,TeV.

None of the selected targets were significantly detected ($>3 \sigma$ significance) in the high-energy $\gamma$-ray range with LAT. In particular, all three sources present a source significance compatible with 0 $\sigma$. In Table\,\ref{tab:gamma} of Appendix\,\ref{sec:appendix_data} we provide the resulting upper limits. 

\subsection{Archival Data}

The radio, infrared, and optical/UV data for {J1407+2827, J1511+0518, and J2022+6137} used for SED modeling were acquired from the NASA/IPAC Extragalactic Database (NED). The relevant references are listed in Table\,\ref{tab:ned} of Appendix\,\ref{sec:appendix_data}. For the data within the near-IR---to---UV range, de-reddening was performed using the extinction law $A_{\lambda} = A_{\rm V} \, a\!(\lambda) + (A_{\rm B} - A_{\rm V}) \, b\!(\lambda)$, with the coefficients $a\!(\lambda)$ and $b\!(\lambda)$ as given in \citet{Cardelli89}, and the B-band and V-band Galactic extinction values $A_{\rm B}$ and $ A_{\rm V}$ as given in the NED following  \citet{Schlafly11}. Source-intrinsic reddening was not taken into account. 

In addition, we also collected de-absorbed fluxes corresponding exclusively to the power-law emission components, obtained via spectral modeling of the broad-band X-ray data collected previously with \chandra, XMM-{\it Newton}, and {\it NuSTAR} by S19 (for {J1407+2827}) and S23 (for J1511+0518 and {J2022+6137}) These power-law components represent the direct intrinsic X-ray continuum emission of our sources.
%\citet[for J1511+0518 and {J2022+6137}]{Sobolewska23}.

\section{Broadband SED Modeling}
\label{sec:model}
    
\subsection{The `Expanding Lobes' Model}

We conducted a modeling analysis of the broad-band SEDs for the three selected CSOs, to probe the origin of their X-ray continuum \emph{power-law} emission. Based on the modeling results, we also investigated the origin of the X-ray reflection components, as well as the apparent $\gamma$-ray quietness of the sources. For this purpose, we explored the model of expanding young radio lobes in AGN, put forward by LS08. In this section, we briefly summarize the model assumptions and model parameters (see Table\,\ref{tab:SEDpar}). We refer the readers to LS08 for further details \citep[see also][further referenced as LO10]{Ostorero10}.
%\citep[see also][]{Ostorero10} .

The model evolves a self-consistent set of equations that describe both the expansion of light relativistic jets and their lobes in a dense ambient medium \citep[see][]{Begelman89}, and the evolution of ultra-relativistic electrons injected into the lobes by the jets, taking into account adiabatic energy losses and radiative cooling. In this framework, a pair of relativistic jets with total kinetic energy $L_{\rm j}$, propagates through the inner segment of the host galaxy, characterized by a constant density $\rho\simeq m_p n_0$. The jets' thrust is balanced by the ram pressure of the ambient medium spread over some area larger than the jet terminal cross-section (due to the effects of a jet precession), so that the jet advance velocity, $v_{\rm h}$, is at most mildly-relativistic. Moreover, the model assumes that all the bulk kinetic energy transported by the jets, is converted at the jet termination shock into the internal energy of the jet particles and the jet magnetic field, amounting together to the lobes' internal pressure $p$. The resulting pressure-driven expansion of the lobes in the direction perpendicular to the jet axis, is supersonic, even though sub-relativistic.

\begin{deluxetable*}{clc}[t]
\label{tab:SEDpar}
\tablecaption{Parameters of the expanding radio lobe model.}
\tablehead{\colhead{Parameter} &\colhead{Description} & \colhead{Values}}
\startdata \hline \\
%& & \\
%\multicolumn{3}{c}{{\bf Parameters measured directly or estimated from the radio and IR data}}\\  
% & & \\
LS  & Radio linear size (half of the hotspot--hotspot distance) &	 Table\,\ref{tab:BB}\\ 
$\nu_{\rm peak}$ & Radio peak frequency (spectral turnover frequency) & Table\,\ref{tab:BB} \\
$v_{\rm h}$ & Jet advance velocity (half of the hotspot--hotspot separation velocity)  & Table\,\ref{tab:BB} \\
 $\nu_{\rm IR}$ & Characteristic frequency of the torus emission & Table\,\ref{tab:BB}  \\
$L_{\rm IR}$ & Infrared luminosity of the circumnuclear torus & Table\,\ref{tab:BB}  \\
 & & \\
 \hline \\
 % & & \\
%\multicolumn{3}{c}{{\bf Model free parameters}}\\  
% & & \\
$n_0$ & ISM number density in the central regions of the host galaxy & Table\,\ref{tab:NHn0} \\ 
$\nu_{\rm UV}$ & Characteristic frequency of the disk UV emission & $2.45 \times 10^{15}$\,Hz {\it (fixed)} \\
$L_{\rm UV}$ & Luminosity of the accretion disk UV emission & Table\,\ref{tab:SEDres} \\ 
$\eta_e$ & Ratio of the radiating electrons' energy density and the total lobes' pressure &  Table\,\ref{tab:SEDres} \\
{$\eta_B$} & Ratio of the magnetic field energy density and the total lobes' pressure & Table\,\ref{tab:SEDres} \\ 
$L_{\rm j}$ & Jet bulk kinetic power &  Table\,\ref{tab:SEDres} \\
 $s_1$ & Low-energy slope of the injected electron population $Q_e\!(\gamma)$ & Table\,\ref{tab:SEDres} \\
$s_2$ & High energy slope of the injected electron population $Q_e\!(\gamma)$ & Table\,\ref{tab:SEDres} \\
$\gamma_{\rm min}$ & Minimum Lorentz factor of the electron energy distribution $Q_e\!(\gamma)$ & 3.0 {\it (fixed)} \\
$\gamma_{\rm br}$ & Break Lorentz factor of the electron energy distribution $Q_e\!(\gamma)$ & Table\,\ref{tab:SEDres} \\
$\gamma_{\rm max}$ & Maximum Lorentz factor of the electron energy distribution $Q_e\!(\gamma)$ & $100 \, m_p/m_e$ {\it (fixed)} \\ 
 & & \\
 \hline
\enddata
\tablecomments{The top part of the table contains parameters measured directly or estimated from the radio and IR data.}
\end{deluxetable*}

Ultra-relativistic electrons and positrons (hereafter referred to as ``electrons'' for simplicity) injected into the lobes, are described by the initial energy distribution $Q(\gamma, t)$, which in our case is assumed to be constant in time and a broken power-law function of electron energy $E_e \equiv \gamma \, m_e c^2$, namely $Q_e\!(\gamma) \propto \gamma^{-s_1}$ for $\gamma_{\rm min} < \gamma < \gamma_{\rm br}$, and $Q_e\!(\gamma) \propto \gamma^{-s_2}$ for $\gamma_{\rm br} < \gamma < \gamma_{\rm max}$. This distribution is evolved in expanding lobes and integrated over the entire source lifetime, taking self-consistently into account adiabatic energy losses, synchrotron emission, and the IC scattering on different photon populations. Those seed photon populations include photons originating from the lobes' synchrotron emission, IR photons originating from the obscuring torus, and UV photons associated with the radiation of the accretion disk. Here we do not consider Comptonization of the starlight emission of host galaxies, noting that this process is of a minor relevance in the case of the youngest and most compact (LS\,$< 100$\,pc) CSOs (see the related discussion in LS08 and LO10). Likewise, the synchrotron self-Compton process is in all the analyzed cases negligible as well.

We assume that the IR and UV spectra of the soft photons are monochromatic, with luminosities ($L_{\rm IR}$ and $L_{\rm UV}$) and frequencies ($\nu_{\rm IR}$ and $\nu_{\rm UV}$) corresponding to the bolometric luminosities and characteristic frequencies of the black bodies which match the observed data points in the respective bands. We note that such an approximation does not significantly affect
the calculated high-energy IC emission of the lobes, given the broad energy range of the lobes' electron energy distribution, $Q_e\!(\gamma)$.

The radio continuum is modeled as synchrotron emission of the lobes, assuming that the low-frequency break in the CSO radio spectrum, around the observed peak frequencies $\nu_{\rm peak}$, is due to the free-free absorption on the interstellar medium (ISM) clouds engulfed by the expanding lobes, and photoionised by the nuclear emission \citep{Begelman99}. 

The lobes' magnetic field and electron energy densities are parameterized as $U_B= \eta_B \, p$ and $U_e = \eta_e \, p$, respectively, where $p$ is the total lobes pressure, and the proportionality factors $\eta_B, \eta_e < 1$ are assumed to be constant during the source evolution (see the discussion in LS08). The linear size of the radio structure, measured from the radio core position to the jet termination shock (``hotspot''), is denoted below as LS.

\subsection{The Modeling Procedure}
\label{sec:modelling}

Among sixteen modeling parameters introduced in the previous section and summarized in Table\,\ref{tab:SEDpar}, three could be measured directly from high-resolution radio monitoring data available for all the analyzed sources. These included radio linear sizes, LS, peak frequencies, $\nu_{\rm peak}$, and jet advance velocities, $v_{\rm h}$, all as given in Table\,\ref{tab:BB} with the corresponding references. Two other modeling parameters, namely the IR torus luminosity $L_{\rm IR}$ and its characteristic frequency $\nu_{\rm IR}$, could be estimated based on the integrated IR data, also as summarized in Table\,\ref{tab:BB}.

We have constrained ranges of the ambient medium density based on the intrinsic equivalent hydrogen column density values, $N_{\rm H}$, provided for each source based on the detailed analysis of the X-ray spectrum (see Table\,\ref{tab:NHn0}). In particular, we estimated lower limits for the ambient medium density utilizing the fact that all three CSOs appeared point-like in the \chandra\ images, and assuming that the X-ray absorbing matter is distributed in a circumnuclear region with a characteristic spatial scale corresponding to $\ell = 1.5^{\prime \prime}$ at the redshift of a source, a typical radius of a \chandra\ source extraction region; this gave $n_{\rm low} \simeq N_{\rm H} / \ell$. Upper limits for the ambient medium density, on the other hand, were found assuming that the characteristic spatial scales of the X-ray absorbers were comparable to the sources' radio linear sizes, namely $n_{\rm high} \simeq N_{\rm H} / {\rm LS}$. Because the range spanned by the $n_{\rm low}$ and $n_{\rm high}$ values is relatively wide, $\sim 2.5$\,dex, we have in addition considered intermediate (in orders of magnitude) densities, $n_{\rm int}$, all as given in Table\,\ref{tab:NHn0}.

\begin{deluxetable}{ccccccccc}
\label{tab:BB}
%\tablecaption{Model parameters measured or estimated from the radio and IR data}
\tablecaption{Parameter values assumed in our modeling, based on the radio and IR data.}
\tablehead{\colhead{Source} & \colhead{LS$^{{\textrm{a}}}$} & \colhead{$\nu_{\rm peak}^{{\textrm{a}}}$} & \colhead{$v_{\rm h}^{{\textrm{b}}}$} &\colhead{$\nu_{\rm IR}^{{\textrm{c}}}$} & \colhead{$L_{\rm IR}^{\textrm{c}}$} \\ 
\colhead{}  & \colhead{pc} & \colhead{GHz} & \colhead{$c$} &\colhead{$10^{13}$\,Hz} & \colhead{$10^{45}$\,erg\,s$^{-1}$} }
\startdata
    {J1407+2827}   &  $9.9$ &$ 4.0$ & $0.13$   & $1.5$ & $0.7$ \\ 
    J1511+0518 &  $7.3$ & $9.8^{\dagger}$ & $0.08$ & $3.0$ & $0.8$ \\
    {J2022+6137}  & $24.5$ & $8.4 $& $0.18$ & $1.0$ &  $1.3$ \\
\hline
\enddata
\tablecomments{{
$^{\textrm{a}}$ References to the high-resolution radio data: \cite{Orienti08,Stanghellini98,Wu13,Tschager00}. \\
$^{\textrm{b}}$ See \cite{An12} and references therein. \\
$^{\textrm{c}}$ For references to the IR data see Table\,\ref{tab:ned}. } \\
$^{\dagger}$ Mean of $\nu_{\rm peak}$ for both lobes.}
\end{deluxetable}

Since the observed IR---to---UV data points were not corrected for the source-intrinsic reddening, the UV emission of an accretion disk was treated as a model free parameter, except that we fixed its comoving characteristic frequency at $\nu_{\rm UV} = 2.45 \times 10^{15}$\,Hz, following LO10.
%However, as the disk UV emission of our sources is expected to be heavily absorbed by circumnuclear dusty tori, and consequently reprocessed and re-radiated in the infra-red band, the intrinsic UV luminosities, $L_{\rm UV}$, should not be lower than the corresponding IR luminosities, $L_{\rm IR}$, constrained with the IR observations. In addition, as the UV disk emission supplies seed photons for the IC scattering that generates X-ray and $\gamma$-ray emissions, the observed levels of the X-ray emission continua, along with the upper limits in the high-energy $\gamma$-ray range, provide constraints on the intrinsic UV luminosities as well.  
Generally, broad-band quasar observations suggest that $L_{\rm IR} \simeq \eta_{\rm t} L_{\rm UV}$, with $\eta_{\rm t} \gtrsim 0.5$ \citep[see in this context, e.g.,][]{Ralowski23}, albeit with a rather wide spread. Therefore, in this work we considered a range of $L_{\rm UV}$ luminosites corresponding to $\eta_{\rm t} = 0.1 - 1$. From this range, we chose the values of $L_{\rm UV}$ that resulted in a good match of the high-energy spectra with the given X-ray and $\gamma$-ray observational constraints.

\begin{deluxetable}{ccccc}
\label{tab:NHn0}
\tablecaption{Adopted range of the ambient medium density, $n_0$.}
\tablehead{\colhead{Source} & \colhead{$N_{\rm H}$} & \colhead{$n_{\rm low}$} & \colhead{$n_{\rm int}$} & \colhead{$n_{\rm high}$}\\ 
\colhead{}  & \colhead{$10^{23}$\,cm$^{-2}$}  & \colhead{cm$^{-3}$}  & \colhead{cm$^{-3}$}  & \colhead{cm$^{-3}$} }
\startdata
    {J1407+2827}   &  4.4 & $70$ & $2,000$ & $20,000$ \\
    J1511+0518 & 0.9  &  $15$ & $400$& $4,000$ \\
    {J2022+6137}  & 3.5 &$25$ & $200$ & $4,000$\\
    \hline
\enddata
\tablecomments{References for $N_{\rm H}$ estimates: S19, S23. }
\end{deluxetable} 

\begin{deluxetable*}{cccccccc}[t!]
\label{tab:SEDres}
\tablecaption{Modeling results.}
\tablehead{\colhead{Source} & \colhead{$U_e/U_B $}  & \colhead{$n_0$}  & \colhead{$L_{\rm UV}$} &\colhead{$L_{\rm j}$} &\colhead{$s_1$} &\colhead{$s_2$} &\colhead{$\gamma_{\rm br}$}  \\ 
\colhead{} & \colhead{} & \colhead{} & \colhead{$10^{45}$\,erg\,s$^{-1}$} &\colhead{$10^{42}$\,erg\,s$^{-1}$} &\colhead{} &\colhead{} &\colhead{$m_p/m_e$} }
\startdata \hline
 & & & & & & & \\
 {J1407+2827}  & 1 & $n_{\rm low} - n_{\rm high}$ & $0.7 - 7.0$ & $17.5 - 5.0 $ & $1.6$ & $5.8$ & $1.0 - 0.8$ \\
 & 10 & $n_{\rm int} - n_{\rm high}$ & $0.7 - 7.0$ & $4.0 - 5.0 $ & $1.6$ & $5.8$ & $0.8$ \\
 & & & & & & & \\
J1511+0518 & $1$ & $n_{\rm int} - n_{\rm high}$ & $1.0 - 3.5$ & $8.0 - 6.0$ & $1.6$ & $5.2$ & $1$  \\ 
 & &  & & & & & \\
{J2022+6137}  & 1 & $n_{\rm low} - n_{\rm high}$ & $13$ & $200 - 40 $ & $1.3$ & $4.7-4.8$ & $1.0$ \\
 & 10 & $n_{\rm low} - n_{\rm high}$ & $1.3 - 13$ & $100 - 27 $ & $1.3$ & $4.7$ & $1.0$ \\
 & & & & & & & \\
 \hline
\enddata
\end{deluxetable*}

Regarding the pressure content of the lobes, in our modeling we investigated two sets of the $\eta_B$ and $\eta_e$ parameters. The first set, $\eta_B = 0.3$ and $\eta_e = 3.0$, corresponds to the scenario in which ultrarelativistic electrons account for the bulk of the lobes' pressure, $p_e \equiv U_e/3 \sim p$, where p stands for the total pressure, while the electron---to---magnetic field energy density ratio is of the order of ten, $U_e/U_B \simeq 10$ \citep[a typical value emerging from modeling relativistic jets in blazar sources; e.g.,][]{Ghisellini10}. The second set, $\eta_B = 0.75$ and $\eta_e = 0.75$, corresponds to the energy equipartition between lobes' ultrarelativistic electrons and magnetic field, $U_e/U_B \simeq 1$, in which case $p_e \sim p/4$. 

For a given combination of $n_0$, $L_{\rm UV}$, $\eta_e$, and $\eta_B$, the model synchrotron (radio) and the IC (X-ray and $\gamma$-ray) emission continua are set by the jet bulk kinetic power, $L_{\rm j}$, and the spectral shape of the injection function, $Q_e\!(\gamma)$. The $L_{\rm j}$ parameter predominantly controls the normalization of the emission continua. However, it also affects the spectral shape. For example, for a given source size, LS, and advance velocity, $v_{\rm h}$, the larger the $L_{\rm j}$ parameter the higher the energy density of the lobes' magnetic field, which leads to an enhanced synchrotron cooling and therefore to softer high-energy spectra of the evolved electron population. Thus, in our modeling procedure, for each considered combination of the $n_0$, $L_{\rm UV}$, $\eta_e$, and $\eta_B$ parameters, we adjust the corresponding $L_{\rm j}$ and $Q_e\!(\gamma)$ values until we obtain a satisfactory match to  the radio data points and, at the same time, the X-ray power-law continua, paying attention also to the upper limits in the high-energy $\gamma$-ray frequency range. 

To reduce the number of the free model parameters, we fixed the minimum and maximum electron Lorentz factors as $\gamma_{\rm min} = 3.0$, and $\gamma_{\rm max} = 100 \, m_p/m_e$, respectively. This choice does not affect the model spectra as long as $s_1<2.0$ and $s_2\gg 2.0$, which in fact is the case for all the three targets (see Table\,\ref{tab:SEDpar} for parameters description).

\section{Results of SED Modeling}
\label{sec:results}

\begin{figure*}[th!]
    \centering
   \includegraphics[width=0.94\textwidth]{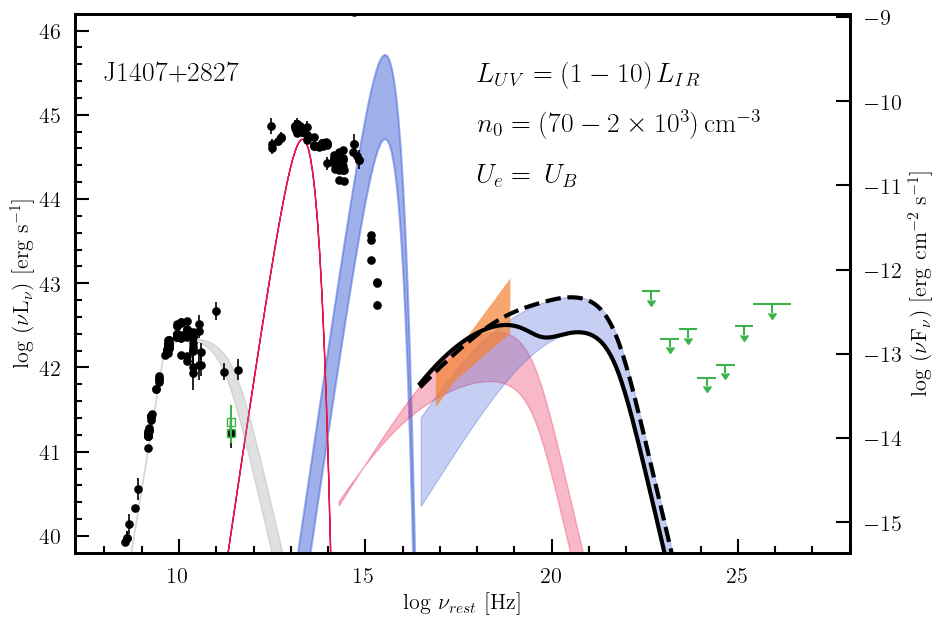}
    \includegraphics[width=0.94\textwidth]{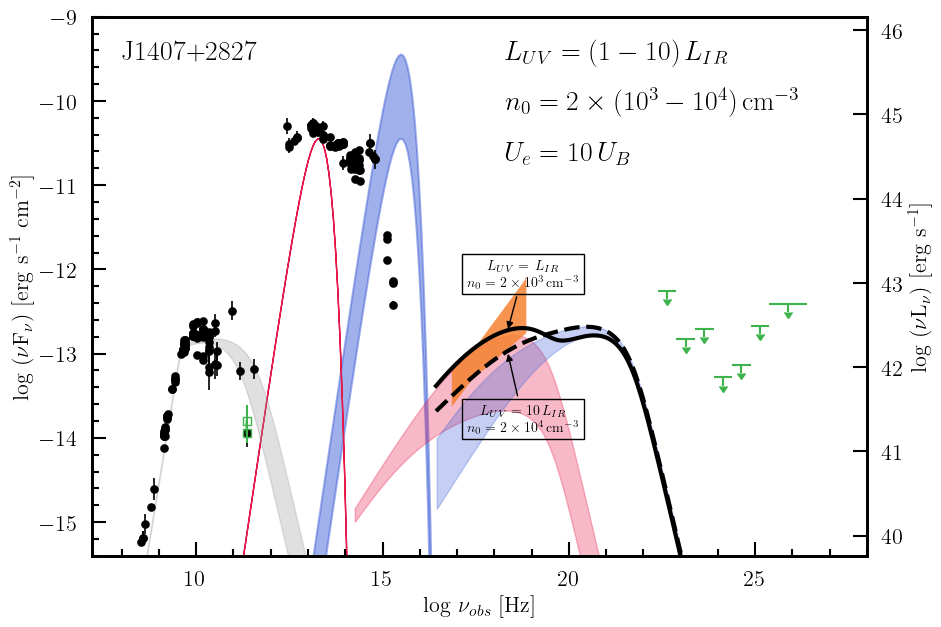}
   \caption{The SED and the best matching models for J1407+2827, corresponding to the model free parameter values/ranges as summarized in Table\,\ref{tab:SEDres} and the figure legend. The archival radio and radio---to---UV data are denoted in the figure by black circles. The new SMA data are plotted as open green squares. The orange bow-tie represents the intrinsic de-absorbed X-ray power-law constraints. Finally, the green arrows illustrate the \fermi-LAT upper limits. The model components include: the lobes' synchrotron emission (gray shaded area), the torus IR emissions (red curve), the disk UV emission (dark-blue shaded area), the IC emission off the disk UV photons (light-blue shaded area), and the IC emission off the torus IR photons (light-red shaded area). Upper panel: radio lobes in equipartition, lower panel: $U_e=10U_B$.}
    \label{fig:SED_OQ208}
\end{figure*}

\begin{figure*}[th!]
    \centering
    \includegraphics[width=0.95\textwidth]{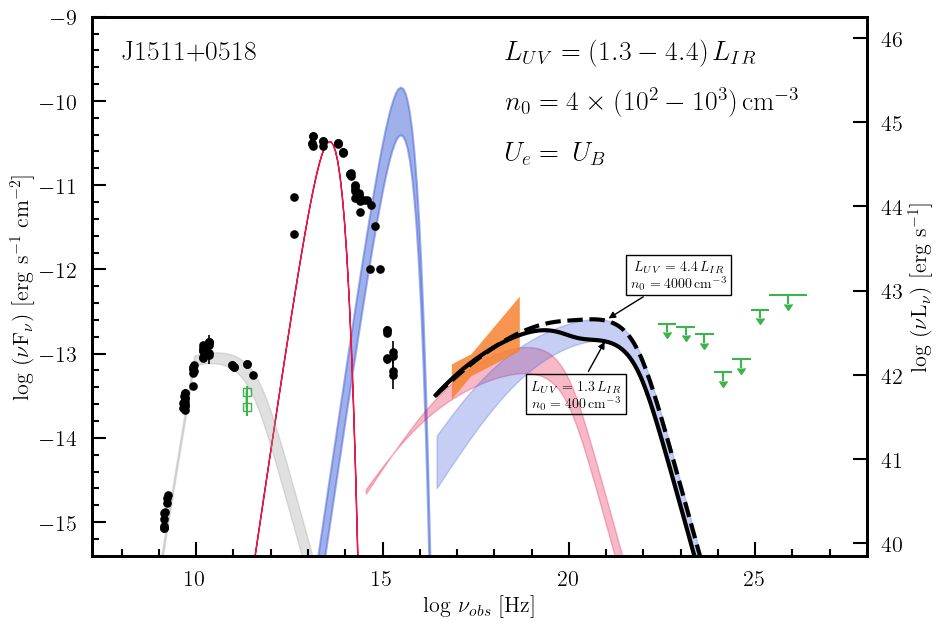}
   \caption{ 
   The SED and the best matching models for J1511+0518 with radio lobes in equipartition. Data and model components marked the same as in Figure\,\ref{fig:SED_OQ208}.} %The thick black solid (dashed) curve denotes the total lobes' IC emission corresponding to the lowest (highest) $n_0$ and $L_{UV}$ from the range indicated in the figure legend.}
    \label{fig:SED_J1511}
\end{figure*}

\begin{figure*}[th!]
    \centering
    \includegraphics[width=0.95\textwidth]{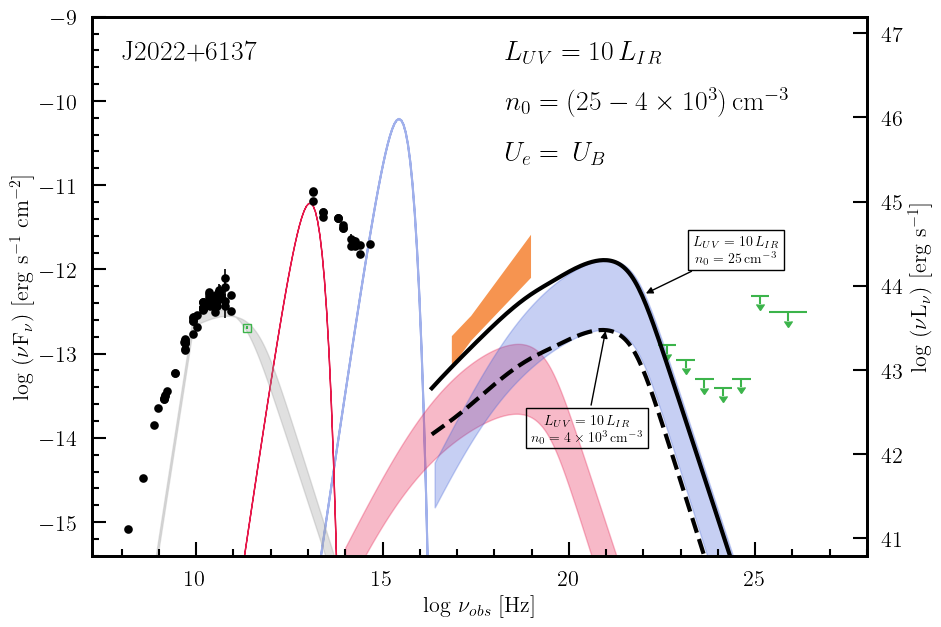}
    \includegraphics[width=0.95\textwidth]{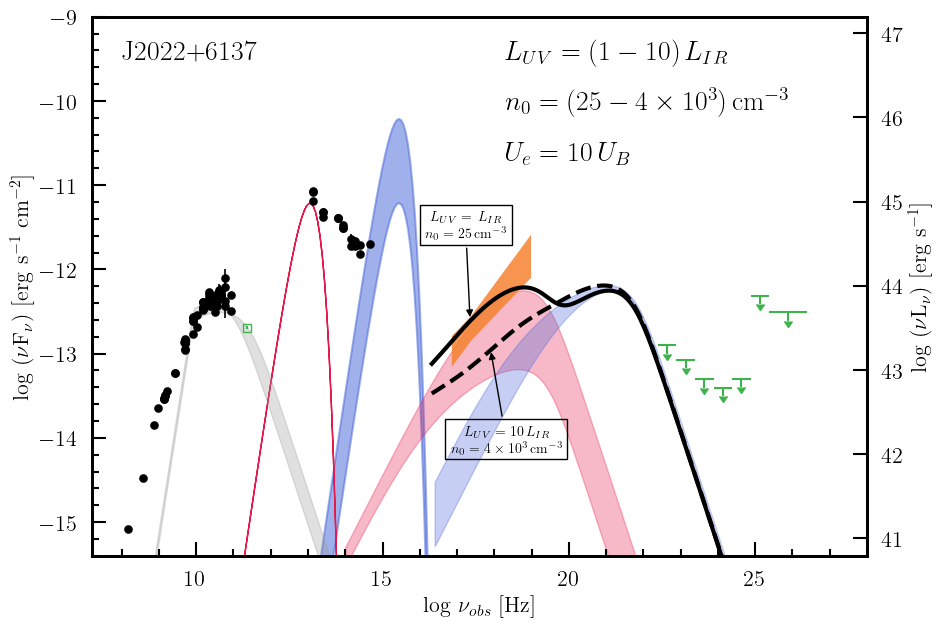}
   \caption{The SED and the best matching models for {J2022+6137}. Upper panel: radio lobes in equipartition, lower panel: $U_e=10U_B$. Data and model components marked the same as in Figure\,\ref{fig:SED_J1511}.}
   \label{fig:SED_J2021}
\end{figure*}

We successfully employed the expanding radio lobe emission scenario to  model the broad-band SED of three CSOs:  {J1407+2827, J1511+0518, and J2022+6137}. For the relevant parameter space of the model we employ, the X-ray emission continuum is produced through the IC scattering of both the torus (IR) and disk (UV) photons, while the $\gamma$-ray emission is generated exclusively through the IC scattering of the UV emission. Since the torus luminosity is constrained by the observed fluxes at IR wavelengths, for a given set of model parameters, a particular $L_{\rm IR}$ value sets a strict lower limit on the X-ray emission of the lobes. The disk luminosity, on the other hand, is a free parameter of the model. For large enough values of $L_{\rm UV}$, the IC scattering of this photon field may dominate over the IC scattering of the torus IR photons, increasing the overall level of the lobes' X-ray continuum. Yet, since high-energy $\gamma$-rays are produced via the IC scattering of the UV photons, the \fermi-LAT upper limits set in turn an upper limit on the allowed value of $L_{\rm UV}$. The high-energy emission output is additionally moderated by the ISM density and the $U_B/U_e$ ratio. Therefore, the X-ray and $\gamma$-ray emission is controlled by the interplay between $L_{UV}$, $n_0$ and $U_B/U_e$ allowing different sets of their values reproduce observed X-ray radiation and follow the $\gamma$-ray flux upper limits.

The free model parameters related to the spectral shape of the electron injection function, $Q_e\!(\gamma)$, are tightly constrained by the observed shape of the radio continua, with the data in the sub-mm range being of a particular importance. Specifically, the shapes of the radio continua require a broken power-law electron injection function with $s_1 < s_2$ with $s_1$ in the $\sim 1.3-1.6$ range. The steep high-frequency segments of the radio continua require that the high-energy slope of the electron energy distribution, $s_2$, is in the range $\sim 4.7-5.8$, that is significantly softer than $s_2 = 2$, which corresponds to a standard diffusive shock acceleration. We revisit this finding in the discussion Section~\ref{sec:discussion} in the context of the $\gamma$-ray quietness of the analyzed targets.

When it comes to the low-frequency segments of the observed radio continua of the discussed CSOs, we note that the applied ``engulfed clouds free-free absorption'' scenario performs extremely well for  {J1407+2827 and J1511+0518}, but fails in reproducing the radio spectrum of {J2022+6137}. In particular, in {J2022+6137} we observe an excess low-frequency radio emission over the model spectra for all combinations of the model free parameters.

\subsection{{J1407+2827}}

In the case of {J1407+2827}, a wide range of the free model parameters under the equipartition condition $U_e/U_B \simeq 1$ could account for the broad-band SED of the source. For sub-equipartition models with $U_e/U_B \simeq 10$ (i.e., $\eta_e = 3.0$ and $\eta_B = 0.3$), the X-ray emission was generally exceeded for any model with the density values below $n_{\rm int}$. The corresponding model solutions are presented in Figure\,\ref{fig:SED_OQ208}, and summarized in Table\,\ref{tab:SEDres}.

%the observed level of the X-ray emission could be produced only within the upper range of the considered ISM density, $n_{\rm int} - n_{\rm high}$. The corresponding model solutions are presented in Figure\,\ref{fig:SED_OQ208}, and summarized in Table\,\ref{tab:SEDres}. 

The broad-band SED of {J1407+2827} was previously discussed in the framework of the expanding radio lobe scenario by LO10. However, they explored only the sub-equipartition models with $U_e/U_B \simeq 10$, and with a relatively low ambient medium density of $n_0 \simeq 0.1$\,cm$^{-3}$. Moreover, the X-ray spectrum considered in LO10, based on the analysis of the initial XMM-{\it Newton} observations \citep{Guainazzi04}, was subject to large uncertainties ($\sim$two orders of magnitude at $\sim0.3$ keV,  one order of magnitude at $\sim10$ keV), preventing robust constraints on the model parameters. In our analysis, we utilized the X-ray continuum and X-ray absorption constraints following from the modeling of an updated broad-band X-ray dataset (S19). As a result, we probed significantly higher ISM density values, $70 - 2\times 10^4$cm$^{-3}$. Finally, we added the new sub-mm and $\gamma$-ray constraints. The main difference between our modeling and the one presented in LO10 was in the jet kinetic luminosity $L_{\rm j}$, which we found to be lower by one to two orders of magnitude, while the parameters of the electron injection function and the soft photon fields were in rough agreement. In particular the value $s_2=5.8$ is 
in agreement with the value reported in LO10 ($s_2=5.6$).

\subsection{J1511+0518}

The SED and best model are shown in Fig.\ref{fig:SED_J1511}. Models that could match the observed SED in both the radio and the X-ray domains, were the equipartition models $U_e/U_B \simeq 1$ (i.e., $\eta_e = \eta_B = 0.75$), with relatively low UV luminosities, $L_{\rm UV} \sim (1.0 - 3.5) \times L_{\rm IR}$, and a relatively high ISM density, within the $n_{\rm int} - n_{\rm high}$ range, as presented in Figure\,\ref{fig:SED_J1511}, and summarized in Table\,\ref{tab:SEDres}. Models with lower ISM densities, $n_0 < n_{\rm int}$, or departing from the equipartition condition, $U_e > U_B$, generally over-predicted the X-ray flux, unless the disk UV luminosity was set to be very low, $L_{\rm UV} \ll L_{\rm IR}$, the possibility which we deem as unlikely.

\subsection{{J2022+6137}}

{J2022+6137} is characterized by an enhanced level of the intrinsic X-ray continuum emission when compared to {J1407+2827 and J1511+0518}. The 0.3-40\,keV luminosity of the source, $\sim 10^{44}$ erg\,s$^{-1}$, could not be obtained within the framework of the applied scenario for high values of the ambient density, regardless of the equipartition ratio $U_e/U_B$, 
without invoking an additional, phenomenological X-ray component.
On the other hand, the ISM density at the level approaching $n_{\rm low}$ and an equipartition parameter $1 \lesssim U_e/U_B \lesssim 10$, allowed us to match the intrinsic X-ray power-law continuum with the lobes' IC emission without violating the \fermi-LAT upper limits. Interestingly, for {J2022+6137}, the $\gamma$-ray upper limits were more restrictive than the X-ray constraints. An increase in $L_{UV}$ or a decrease in $n_0$ violated the $\gamma$-ray limits before the X-ray limits.
%{Only a combination of ISM density at the level approaching $n_{\rm low}$ and an equipartition parameter $U_e/U_B > 1$ allowed us to match the intrinsic X-ray power-law continuum with the lobes' IC emission without violating the \fermi-LAT upper limits.}
%For any other choice of the free model parameters, an additional X-ray component seemed %required.
The results of the modeling for this source are presented in Figure\,\ref{fig:SED_J2021}, and summarized in Table\,\ref{tab:SEDres}.

\section{Discussion and Conclusions}
\label{sec:discussion}

In this work, we studied the broad-band emission of three highly X-ray obscured active galaxies, with a CSO radio classification, {J1407+2827}, J1511+0518, and {J2022+6137}, each featuring young, compact radio lobes with linear sizes up to 25 pc. We demonstrated that the IC scattering of the IR and UV photons (supplied by a dusty torus and an accretion disk, respectively) within the expanding radio lobes, provides a plausible source of intrinsic X-ray continuum emission on the levels consistent with the observational constraints.

\begin{figure}[th!]
    \centering
    \includegraphics[width=\columnwidth]{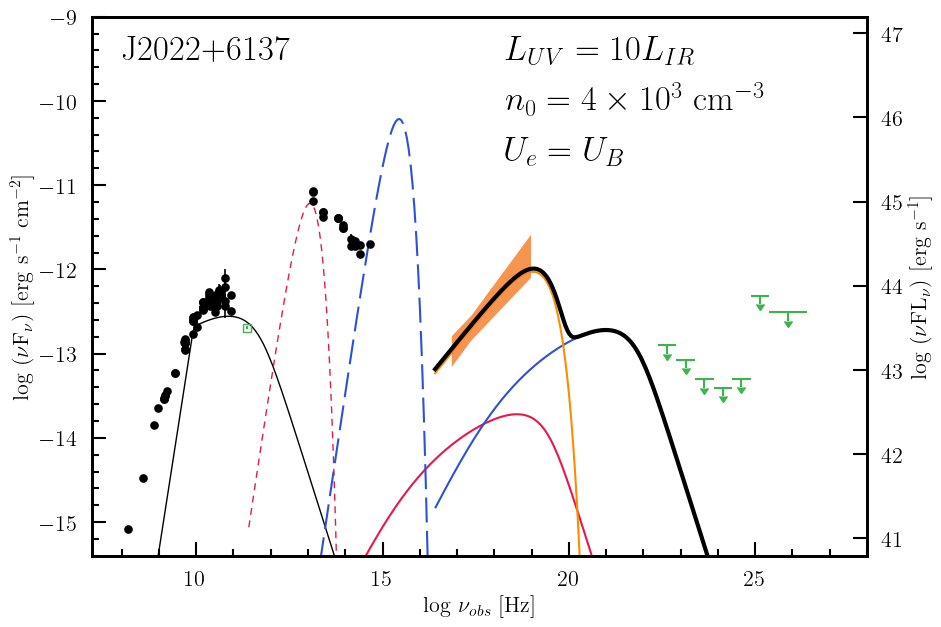}
    \caption{An example of the expanding radio lobe model solution applied to the SED of {J2022+6137}, which underestimates the most the intrinsic X-ray emission (see the plot label and Table~\ref{tab:SEDres} for model parameters). Data, intrinsic X-ray emission and \fermi-LAT upper limits are plotted the same as in Figure\,\ref{fig:SED_J2021}. An additional , phenomenological X-ray component with the photon index of $1.45$ and the $2-10$\,keV luminosity $\sim 9.95 \times 10^{43}$\,erg\,s$^{-1}$ is required (orange line) to match the intrinsic X-ray emission, and it dominates the X-ray band. The model components include: the lobes' synchrotron emission (gray solid line), torus IR emissions (red short-dashed line), disk UV emission (blue long-dashed line), IC emission of the disk seed photons (blue solid line), IC emission of the torus seed photons (red solid line). The thick solid black line corresponds to the sum of the IC components from the lobes and the additional X-ray component.}

    \label{fig:2021_corona}
\end{figure}

For the two analyzed CSOs, {J1407+2827} and J1511+0518, certain models could be ruled out because they overestimated the intrinsic X-ray emission, while none of the probed models were found to underestimate the intrinsic X-ray emission or to violate the $\gamma$-ray upper limits. Models accounting for the comparatively high $0.3-40$\,keV luminosity of {J2022+6137} ($\sim 10^{44}$\,erg\,s$^{-1}$), and at the same time complying with its $\gamma$-ray constraints, were characterized by a low ambient medium density and high ultraviolet luminosity, $L_{\rm UV} = 10 L_{\rm IR}$, in the case of lobes in the equipartition. A range of $n_0$ and $L_{UV}$ values, consistent with the lower end of the range depicted in Figure~\ref{fig:SED_J2021}, was allowed for the case of the lobes in sub-equipartition. Since the high density models underestimated the X-ray emission of {J2022+6137}, we note that an additional, phenomenological X-ray continuum component may be present in the SED of {J2022+6137}. In Figure~\ref{fig:2021_corona} we show an example of a model consisting of the emission of the expanding radio lobes with $n_0 = 4 \times 10^3$\,cm$^{-3}$, and a phenomenological cut-off power-law component with a photon index of $1.45$ (the best-fit photon index of the intrinsic de-absorbed X-ray power-law in S19). The photon index of 1.45 is within the range observed in AGNs and X-ray binaries \citep{Yang15}. This additional component, with the $2-10$\,keV luminosity of $\sim 10^{44}$\,erg\,s$^{-1}$, could originate, e.g., due to a compact lamp-post corona very close to the black hole or due to a complex structure of the jetted outflow (see \citealt{Migliori14} and \citealt{Sobolewska22} in the context of young radio sources).

Apart from its high X-ray luminosity, {J2022+6137} stands out in our sample also because of an excess low-frequency radio emission compared to the model spectrum, for any combination of the free model parameters. This could signal a presence of an additional, extended (on arcsec scales) radio component \citep[see in this context][]{Kharb10}.
It is also possible that the engulfed clouds free-free absorption scenario is not applicable in the case of {J2022+6137}.

\begin{figure*}
    \centering
    \includegraphics[width=0.49\textwidth]{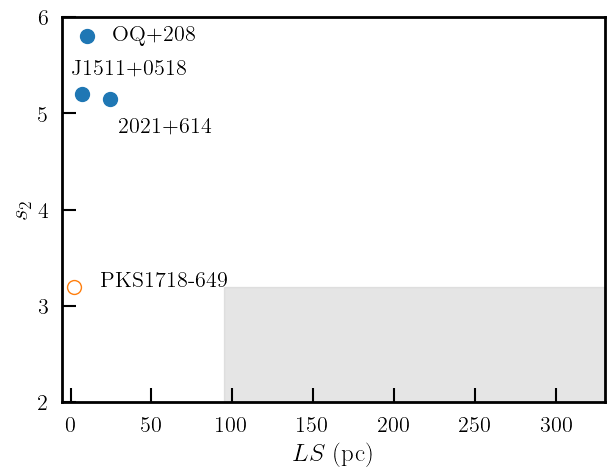}
    \includegraphics[width=0.49\textwidth]{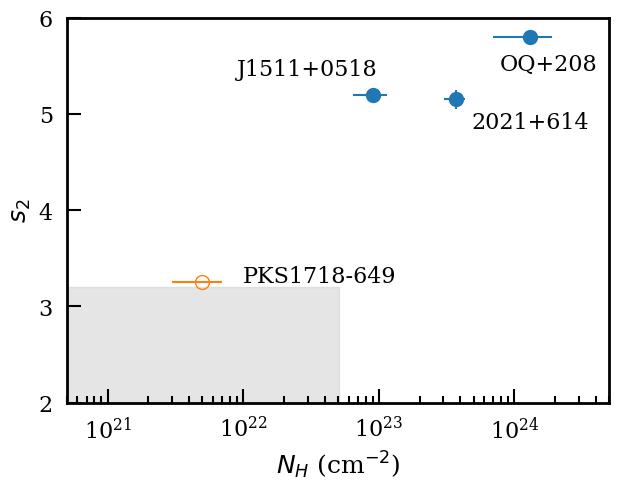}
   \caption{Comparison of the high-energy slope of the injection electron distribution, $s_2$, with the radio linear size LS (left panel), and the source-intrinsic absorbing column density $N_{\rm H}$, for the three obscured CSOs analyzed in this paper, {J1407+2827}, J1511+0518 and {J2022+6137}, the X-ray unobscured PKS\,1718--649 modeled in \cite{Sobolewska22}, and other CSOs with low $N_{\rm H}$ modeled by LO10 and depicted in the figure by the gray shaded areas. }
    \label{fig:s2}
\end{figure*}

For all three sources, we compared the resulting $L_{\rm UV}$ and $L_{\rm j}$ values, to those estimated in AW20: $31\times10^{42}$\,erg\,s$^{-1}$ ({J1407+2827}), $L_{\rm j}= 10.5 \times 10^{42}$\,erg\,s$^{-1}$ (J1511+0518), $220\times10^{42}$\,erg\,s$^{-1}$ ({J2022+6137}); and $L_{\rm UV}=4.3\times10^{45}$\,erg\,s$^{-1}$ ({J1407+2827}) $1.5\times10^{45}$\,erg\,s$^{-1}$ (J1511+0518), $1.5\times10^{45}$\,erg\,s$^{-1}$ ({J2022+6137}). Due to the high absorption of these CSOs, AW20 estimated $L_{UV}$ through the SED for J1511+0518, \citep[following the][method]{Trichas13}, based on the $12$\,$\mu$m and OIII lines respectively for {J1407+2827} and {J2022+6137} \citep{Kosmaczewski20,Wu09}. The inferred $L_{UV}$ values from our analysis are in agreement with those presented by AW20. The jet power resulting from our modeling is also roughly in agreement with the values estimated by AW20. The differences, which are up to a factor of a few, could result from the different assumptions about lobe geometry and the strict energy equipartition assumed in AW20. Note in this context that the emerging total jet power for {J1407+2827}  and J1511+0518 are $\leq 10^{44}$\,erg\,s$^{-1}$, below the threshold for an effective confinement of the lobes within the ISM. Therefore they are expected to be trapped for a relatively long time before they expand outside of the host galaxies \citep{Mukherjee16,Mukherjee17}, whereas the total jet power of {J2022+6137} for some models is above this threshold. 

Modeling the radio continua, including newly acquired SMA data, indicated that the electron energy spectrum has to be of a broken power-law form, with $s_1\leq2$ below $\gamma \sim m_p/m_e$, and $s_2 \gg 2$ at higher energies. \fermi-LAT observations reinforced this conclusion because with no sharp break in the electron energy distribution, $s_2-s_1 > 2$, the current LAT upper limits for the source fluxes would be exceeded. Moreover, we emphasize that this broken power-law form of the energy distribution corresponds to the electron `injection' spectrum. Thus, it described the electron spectrum formed at the acceleration site, namely, the jet termination shock, and not the spectrum modified by the subsequent evolution within the expanding and radiatively cooling lobes. This finding aligns with the scenario proposed by \cite{Stawarz07}, where a broken power-law electron energy distribution with a critical break energy around $m_{\rm p}/m_{\rm e}$ represents two distinct regimes of electron acceleration at mildly-relativistic, proton-dominated shocks with perpendicular magnetic field configurations. The main hypothesis is that in a proton-mediated shock, only electrons with energies exceeding those of cold protons ($\gamma > m_{\rm p}/m_{\rm e}$) can undergo diffusive shock acceleration. This process is expected to yield steep electron spectra (energy indices $\gg 2$) at mildly-relativistic perpendicular shocks. Electrons with lower energies ($\gamma < m_{\rm p}/m_{\rm e}$) are likely accelerated by different mechanisms, possibly in the near upstream of the shock (as discussed and referenced in \citealt{Stawarz07}).

In Figure\,\ref{fig:s2}, we compare the $s_2$ parameter values emerging from our modeling of the three heavily X-ray obscured CSOs, with those obtained for X-ray un-obscured CSOs by LOS10 (excluding {J1407+2827}) and \citet[PKS\,1718--649]{Sobolewska22}, using the same expanding radio lobe model. We plot $s_2$ as a function of the projected linear radio size, LS, and as a function of the source-intrinsic absorbing column density, $N_{\rm H}$. It can be seen that the $s_2$ index correlates with $N_{\rm H}$, and that the most compact CSOs, with LS $\lesssim 25$ pc, have softer $s_2$ indices than CSOs with LS $\gtrsim 100$ pc.

In our modeling the absorbing column density guides the values of the ambient medium density into which the newly born CSO jets propagate, $N_{\rm H} \propto n_0$, the $s_2 - N_{\rm H}$ correlation observed in our small sample agrees very well with the $s_2 - n_0$ trend reported in the model data for the evolved radio galaxies by \citet{Wojtowicz21}. This allows us to draw the same conclusion as \citeauthor{Wojtowicz21}, namely that the efficiency of the electron acceleration at mildly-relativistic termination shocks of AGNs jets decreases with increasing density of the ambient medium. Moreover, very soft values of $s_2$ may explain the apparent quietness of compact and heavily obscured CSOs in the $\gamma$-ray range, since it is the high-energy slope of the electron energy distribution that effectively suppresses the lobes' IC emission at GeV photon energies for a given total jet kinetic power and a given disk UV luminosity \citep[compare to the case of the $\gamma$-ray detected PKS\,1718--649 discussed in][]{Sobolewska22}.

\begin{figure*}
    \centering
    \includegraphics[width=0.49\textwidth]{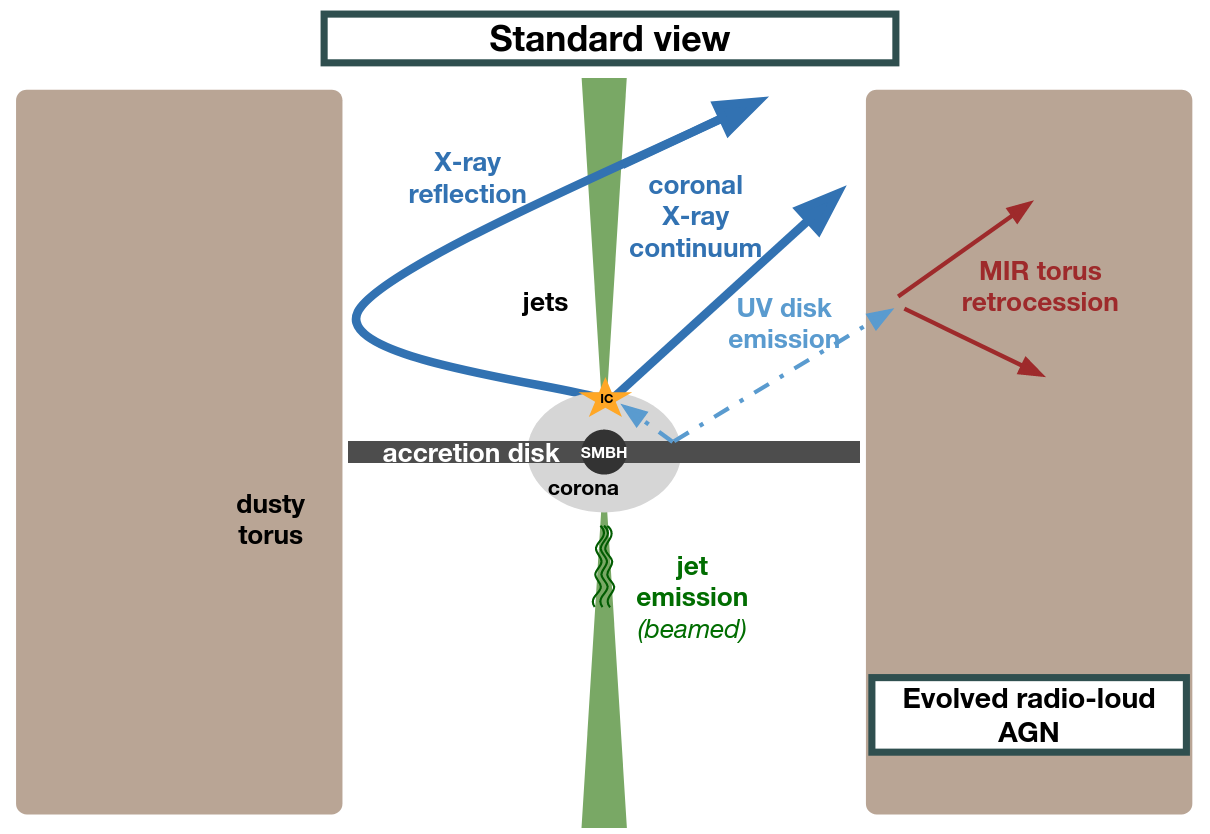} 
    \includegraphics[width=0.49\textwidth]{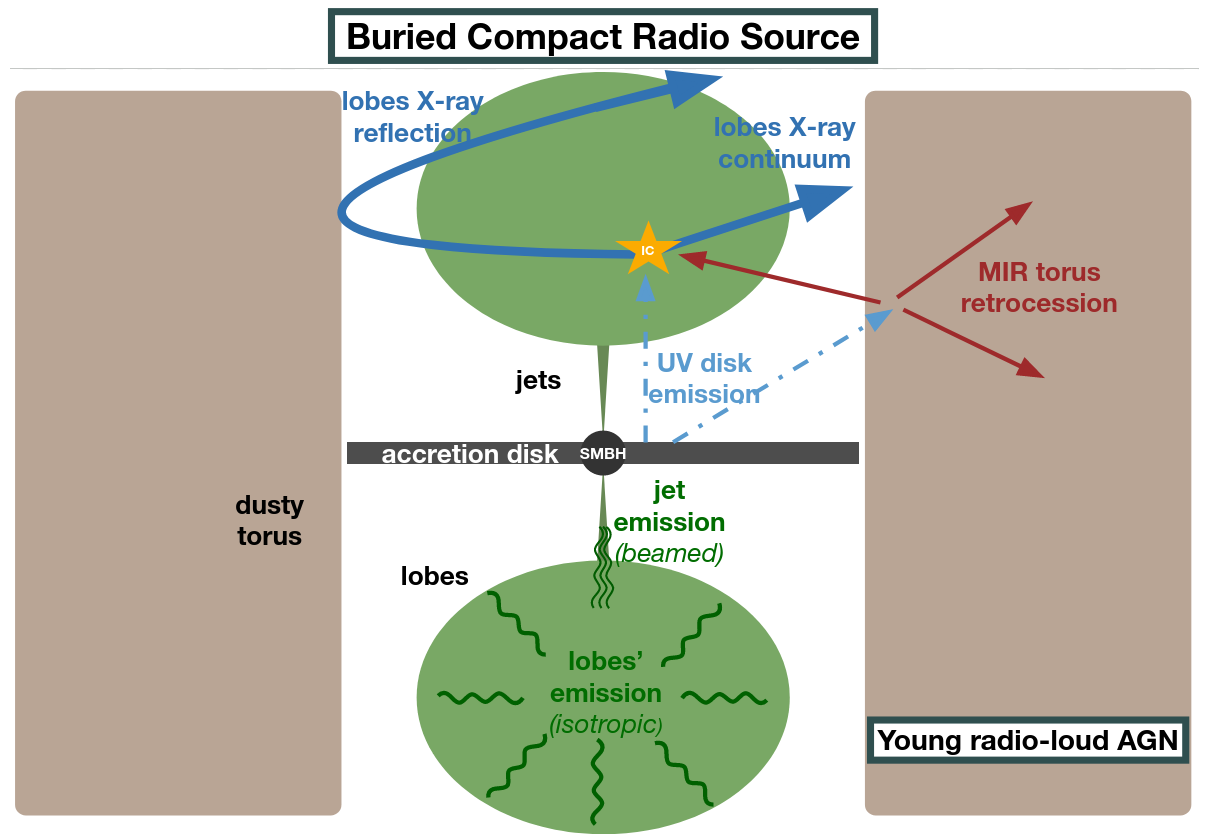} 
    \caption{{\it Left panel:} Schematic view of an evolved radio-loud AGN, depicting the UV disk emission comptonized within ultra-compact disk hard X-ray emitting region, and subsequently reflected by a dusty torus to form the reflection X-ray component. {\it Right panel:} Schematic view of a young radio source, where the X-ray continuum emission is produced within compact radio lobes, formed by newly-born relativistic jets propagating through the circumnuclear environment.}
    \label{fig:cartoon}
\end{figure*}
The ISM number densities resulting from our modeling are relatively high, orders of magnitude higher than the fiducial value of $n_0 = 0.1$\,cm$^{-3}$ considered in the analogous modeling of CSOs by LO10. However, all the sources analyzed in LO10, except {J1407+2827}, have linear radio sizes exceeding 100\,pc, and one should expect lower ambient medium densities for them when compared to the three particularly compact/young targets discussed here. Indeed, assuming a general scaling of $n_0$ with the square of the distance from the black hole, as appropriate for the winds from black hole accretion flows \citep[see, e.g.,][]{Cui20}, gas number densities at the level of $\sim 10^3$\,cm$^{-3}$ at 10\,pc scale, would correspond to $\sim 0.1$\,cm$^{-3}$ at the distance of 1\,kpc. 

We can also compare our estimates of the ISM density with the estimates obtained for galaxies with a radio-loud tidal disruption event (TDE). \cite{Cendes23} presented a collection of ISM density profiles in TDE hosts, constructed assuming an external shock model (see however in this context \citealt{Pasham18}). For the distance scale corresponding to $1-25$ pc (linear radio size of the lobes in our sources), expressed in Schwarzschild radii \citep[$10^5-10^6$ Schwarzschild radii for black hole masses $\sim 10^8-10^9$\,M$_{\odot}$;][]{Wojtowicz20}, the ISM density estimates of \citeauthor{Cendes23} cover a density range, $\sim 10 - 10^3$\,cm$^{-3}$, similar to that resulting from our work.

X-ray reflection continua, including the fluorescent narrow Iron K$\alpha$ line at 6.4\,keV, have been detected in all three CSOs (S19, S23). This raises a question regarding the origin of the X-ray reflection component in these objects. Interestingly, these galaxies fall within the Quasar/Seyfert region on the WISE color diagram and their mid-IR colors align with the dominant emission from circumnuclear dust \citep{Kosmaczewski20,Nascimento22}. Additionally, the linear sizes of their radio lobes are of the same order as those of dusty tori in numerous Seyfert Galaxies, as measured by ALMA \citep[e.g.,][]{Combes19,Garcia21}. Therefore, we propose that in all three cases, the reflection components may in fact be related to the isotropic IC emission from the lobes reflecting off the tori.

In Appendix\,\ref{sec:appendix} we provide the energetic considerations, concluding that the lobes' high-energy (X-ray and $\gamma$-ray) emission, may indeed dominate over the disk corona radiative output in the case of the youngest and the most powerful CSOs. We emphasize in this context that, in both cases, we are dealing with the accretion power (or, to be more precise, a combination of the accretion power and the black hole rotational energy), but, nonetheless, the mechanism for extracting a fraction of this power to produce the high-energy emission is quite distinct. In particular, the energy radiated away in the high-energy domain is either at the expense of a disk magnetic energy dissipated within disk corona to hot electrons \citep[see, e.g.,][]{Sridhar21}, or a bulk kinetic energy of relativistic jets converted at the jet termination shock to the internal energy of the lobes' ultra-relativistic electrons.

In Figure\,\ref{fig:cartoon}, we present a sketch illustrating the proposed scenario. In the left panel of the figure, we present a schematic ``standard'' view of an evolved radio-loud AGN, depicting in particular how the UV disk emission is IC scattered within an ultra-compact region filled with hot electrons (e.g. a lamp-post corona above the accretion disk, a hot inner flow, or a base of a jet), giving rise to the X-ray continuum emission, which is subsequently reflected from a dusty torus to form the X-ray reflection component.
The right panel of the figure illustrates a young radio source, where the X-ray continuum emission originates within the compact radio lobes, inflated by the newly-born relativistic jets propagating through the circumnuclear environment. Such a scenario predicts that the X-ray emission of CSOs whose radio lobes expanded beyond the scale corresponding to the size of a dusty torus should not show strong absorption and reflection features. So far, the existing X-ray observations of the CSO sources support this idea in that they have not identified X-ray absorbed CSOs or CSOs showing a 6.4\,keV Iron fluorescence among the sources with radio sizes $\gtrsim 40-50$\,pc \citep[e.g.,][]{Sobolewska19a}.

Interestingly, due to the rather different geometries of the systems shown in the left and right panels of Figure\,\ref{fig:cartoon}, as well as different spatial scales and energies of the radiating electrons involved, variability patterns and polarization properties of the X-ray continuum emission produced in the framework of the ``standard'' model and our CSO scenario should be quite distinct. In particular, in the CSO scenario we expect a rather steady X-ray continuum (power-law) emission, involving possibly only a gradual monotonic decrease on the timescale of years, as opposed to a stochastic red-noise variability expected in the case of an ultra-compact X-ray source \citep{Uttley14}. An in-depth spectral modeling of the reflection component in the framework of our model, along with an analysis of the X-ray polarization properties, keeping in mind the capabilities of current and future X-ray polarimetry missions, will be presented in a forthcoming paper.
 
\section{Summary}
\label{sec:summary}
We studied the broadband spectral energy distribution of three galaxies with CSO radio classification: {J1407+2827}, J1511+0518, and {J2022+6137}. 
The radio morphologies of these galaxies are characterized by small projected linear sizes (up to $25$ pc). Additionally, in X-rays they are heavily absorbed, $N_{\rm{H}}>10^{23}$\,cm$^{-2}$, and display fluorescent Fe\,K$\alpha$ lines in their spectra. The main findings from our modeling of their radio to $\gamma$-ray emission are summarized as follows:
\begin{itemize}
    \item[1)]  The high-energy emission of these three sources can be explained within the framework of an expanding radio lobe model. Namely, the X-ray continuum originates from radio lobes that are still embedded within the dusty tori. These lobes constitute a reservoir of relativistic electrons producing high-energy emission through Comptonization. This emission is then reflected from the torus, adding fluorescent Fe\,K$\alpha$  lines and the reflection component.
    \item[2)] The proposed scenario may result in different polarization properties than the polarization from  a ``standard'' compact corona. These properties will be considered in future work.
    \item[3)] Given the energetics of the system, detailed in Appendix~\ref{sec:appendix}, the high-energy emission of the most compact radio lobes might dominate over the disk-corona radiative output. 
    \item[4)] The high-energy slope of the broken power-law electrons' energy distribution at injection has to be notably soft ($s_2\simeq4.7-5.8$) to account for the spectral shape of the radio emission in all three sources. Relative to other CSOs, this index appears to correlate with the X-ray absorbing column density (the higher $N_{\rm{H}}$, the higher $s_2$) and to anti-correlate with source size (the larger the object, the smaller $s_2$). {The softness of the high-energy index of the electron energy spectrum at the injection suggests that the efficiency of particle acceleration decreases with the density of an ambient medium. This anti-correlation between acceleration efficiency and n$_0$ might be responsible for the $\gamma$-ray quietness of these sources}
    %The soft high-energy index of the electron energy spectrum at the injection suggests that the efficiency of particle acceleration decreases with the density of an ambient medium. 
    Additionally, it might be responsible for the $\gamma$-ray quietness of these sources. {In general, the obtained ISM density values are rather high, ranging from  $\sim 10$ to even $10^4$\,cm$^{-3}$. Interestingly, the ISM densities obtained through our modeling agree with estimates obtained from the multi-epoch radio follow-ups of tidal disruption events in several galaxies, at distances corresponding to the radio lobes size in our sources.}
    \item[5)] The broadband spectral properties of {J2022+6137} differ from those of {J1407+2827} and J1511+0518 in several aspects.
    The ratio of the intrinsic (de-absorbed) X-ray power-law flux to the upper limits on the $\gamma$-ray flux is significantly higher in {J2022+6137} than in the case of the other two CSOs. Consequently, for some models, the data of J2021+624 require an additional X-ray emission component \citep[as in the case of PKS 1718-649;][]{Sobolewska22}. This additional component can be associated with e.g., a compact corona, a base of a jet, or a hot inner flow.
    The emerging total jet power of {J1407+2827} and J1511+0518 is $\lesssim 10^{44}$\,erg\,s$^{-1}$. However, in {J2022+6137} the jet power of the lowest density models reaches and even exceeds this threshold, which suggests that jet expansion beyond the host galaxy is more likely in the case of {J2022+6137} than in the case of other two considered sources \citep{Mukherjee16,Mukherjee17}.
    Finally, our model under-predicts the radio emission of {J2022+6137} below the spectral break, signaling either the presence of an additional, extended radio component or indicating that it cannot be described within the framework of enfluged clouds free-free absorption.  
\end{itemize}

\begin{acknowledgments}
\newpage
D.~\L{}.~K. and M.~S. acknowledge NASA \chandra\ contract GO2-23110X.
M.~S. and A.~S. was supported by NASA contract NAS8-03060 (Chandra X-ray Center).
D.~{\L}.~K. and {\L}.~S. were supported by the Polish National Science Center grants 2016/22/E/ST9/00061 and DEC-2019/35/O/ST9/04054. G.~P. acknowledges support by ICSC – Centro Nazionale di Ricerca in High Performance Computing, Big Data and Quantum Computing, funded by European Union – NextGenerationEU. 

{This paper employs a list of Chandra datasets, obtained by the Chandra X-ray Observatory, contained in~\dataset[DOI: https://doi.org/10.25574/cdc.213]{https://doi.org/10.25574/cdc.213}.}

The SMA is a joint project between the Smithsonian Astrophysical Observatory and the Academia Sinica Institute of Astronomy and Astrophysics and is funded by the Smithsonian Institution and the Academia Sinica. We recognize that Maunakea is a culturally important site for the indigenous Hawaiian people; we are privileged to study the cosmos from its summit.

The Fermi LAT Collaboration acknowledges generous ongoing support from a number of agencies and institutes that have supported both the development and the operation of the LAT as well as scientific data analysis. These include the National Aeronautics and Space Administration and the Department of Energy in the United States, the Commissariat à l'Energie Atomique and the Centre National de la Recherche Scientifique / Institut National de Physique Nucléaire et de Physique des Particules in France, the Agenzia Spaziale Italiana and the Istituto Nazionale di Fisica Nucleare in Italy, the Ministry of Education, Culture, Sports, Science and Technology (MEXT), High Energy Accelerator Research Organization (KEK) and Japan Aerospace Exploration Agency (JAXA) in Japan, and the K. A. Wallenberg Foundation, the Swedish Research Council and the Swedish National Space Board in Sweden. Additional support for science analysis during the operations phase is gratefully acknowledged from the Istituto Nazionale di Astrofisica in Italy and the Centre National d'Etudes Spatiales in France. This work is performed in part under DOE Contract DE-AC02-76SF00515.

\end{acknowledgments}

\vspace{5mm}

\software{CIAO v.4.14 \citep{Fruscione06},
Sherpa \citep{Freeman01}
          }

\bibliographystyle{aasjournal}

\appendix
\restartappendixnumbering
\section{Observational data}
\label{sec:appendix_data}
Here we present the observational data used in our analysis. The \fermi-LAT upper limits on the CSOs fluxes within seven narrow energy bins are listed in Table\,\ref{tab:gamma}. The SMA data are presented in Table\,\ref{tab:SMA}. For details on the data processing see sec. \ref{subsec:SMA} -- \ref{subsec:fermi}. The list of relevant references for the multi-wavelength data acquired from the NASA/IPAC Extragalactic Database (NED) is provided in Table\,\ref{tab:ned}.  
\begin{deluxetable}{ccc}[h]
\label{tab:gamma}
\tabletypesize{\footnotesize}
\tablecaption{\fermi-LAT upper limits.}
\tablehead{\colhead{Object} & \colhead{Energy band} & \colhead{Upper limit} \\
\colhead{} & \colhead{GeV} & \colhead{$10^{-13}$\,erg cm$^{-2}$s$^{-1}$}}
\startdata
{}{J1407+2827}     &  0.100 -- 0.316  &	$5.53$  \\ 
          &  0.316 -- 1.000  & $1.51$    \\
          &  1.000 -- 3.162  & $1.97$ \\ 
          & 3.162 -- 10.00 &$0.52$  \\ 
          &  10.00 -- 31.62 & $0.74$ \\ 
          & 31.62 -- 100.0 & $2.14$ \\
          & 100.0 -- 1,000 & $3.88$ \\ \hline
{J1511+0518}     &  0.100 -- 0.316  &	$2.26$  \\ 
          &  0.316 -- 1.000  & $2.07$    \\
          &  1.000 -- 3.162  & $1.69$ \\ 
          & 3.162 -- 10.00 &$0.60$  \\ 
          &  10.00 -- 31.62 & $0.86$ \\ 
          & 31.62 -- 100.0 & $3.33$ \\
          & 100.0 -- 1,000 & $4.92$ \\ \hline
{J2022+6137}     &  0.100 -- 0.316  &	$1.25$  \\ 
          &  0.316 -- 1.000  & $0.85$    \\
          &  1.000 -- 3.162  & $0.49$ \\ 
          & 3.162 -- 10.00 &  $0.39$  \\ 
          &  10.00 -- 31.62 & $0.50$ \\ 
          & 31.62 -- 100.0 & $4.89$ \\
          & 100.0 -- 1,000 & $3.11$ \\
\hline
\enddata
\end{deluxetable}

\begin{deluxetable*}{cclllcc}
\label{tab:SMA}
\tablecaption{Details of the SMA observations.}
\tablehead{\colhead{Source}  & \colhead{Redshift} & \colhead{RA} & \colhead{Dec} & \colhead{Date} & \colhead{$\nu$}& \colhead{Flux} \\ \colhead{}  & \colhead{} & \colhead{J2000} & \colhead{J2000} & \colhead{} & \colhead{GHz} & \colhead{mJy} }
\startdata    
{J1407+2827}    & 0.076 & 14h\,07m\,0.4s & +28$^{\circ}$\,27$^{\prime}$\,14.7$^{\prime \prime}$ &08-Aug-2018 & 227.225& $6.96 \pm  3.20$  \\ 
    J1511+0518 & 0.084 & 15h\,11m\,41.3s &+05$^{\circ}$\,18$^{\prime}$\,09.3$^{\prime \prime}$ & 13-Jul-2018 & 225.561 & $10.22\pm 2.55$   \\
              & & & & 08-Aug-2018 &225.558 & $15.46 \pm 2.98$   \\
              & & & 	 &02-May-2019& 225.558 &$ 5.01 \pm 1.35$   \\
    {J2022+6137}  & 0.227 & 20h\,22m\,06.7s  &+61$^{\circ}$\,36$^{\prime}$\,58.8$^{\prime \prime}$ & 13-Jul-2018& 225.561 &$90.11 \pm  2.89$       \\
\hline
\enddata
\end{deluxetable*} 

\begin{deluxetable*}{p{0.11\linewidth}p{0.11\linewidth}p{0.71\linewidth}}
\tablecaption{References to the archival radio, infrared, and optical/UV data.}
\label{tab:ned}
\tablehead{\colhead{Energy band} &\colhead{Source} & \colhead{References}}
\startdata
{ Radio} &{J1407+2827} & \cite{OQ208_RADIO1}, \citet{OQ208_RADIO2}, \citet{J1511_Radio11}, \citet{OQ208_RADIO4}, \cite{OQ208_RADIO5}, \citet{OQ208_RADIO6}, \citet{J1511_Radio1}, \citet{OQ208_RADIO8}, \cite{OQ208_RADIO9}, \citet{OQ208_RADIO10}, \citet{OQ208_RADIO11}, \citet{OQ208_RADIO12} \cite{OQ208_RADIO13}, \citet{OQ208_RADIO14}, \citet{J1511_Radio6}, \citet{OQ208_RADIO16} \cite{OQ208_RADIO17}, \citet{J1511_Radio9}, \citet{OQ208_RADIO19}, \citet{OQ208_RADIO20} \cite{J1511_Radio3}, \citet{J1511_Radio13}, \citet{OQ208_RADIO23}, \citet{OQ208_RADIO24} \cite{OQ208_RADIO25}, \citet{OQ208_RADIO26}, \citet{OQ208_RADIO27}, \citet{J1511_Radio8}, \cite{J1511_Radio7}, \citet{OQ208_RADIO30}, \citet{Stanghellini98}, \citet{OQ208_RADIO32} \cite{J1511_Radio2}, \citet{OQ208_RADIO34}, \citet{OQ208_RADIO35}, \citet{OQ208_RADIO36} \cite{OQ208_RADIO37}, \citet{J1511_Radio10} \\
& {J1511+0518} & \cite{J1511_Radio1}, \citet{J1511_Radio2}, \citet{J1511_Radio3},  \citet{J1511_Radio5}, \citet{J1511_Radio6}, \citet{J1511_Radio7}, \citet{J1511_Radio8} \cite{J1511_Radio9}, \citet{J1511_Radio10}, \cite{J1511_Radio11}, \citet{J1511_Radio12}, \citet{J1511_Radio13} \citep{J1511_Radio14}, \citet{J1511_Radio15}, \citet{J1511_Radio16}, \citet{J1511_Radio17} \\
 & {J2022+6137} & \cite{J2021_RADIO1}, \citet{J1511_Radio1}, \citet{J2021_IR7}, \citet{J2021_RADIO4}, \cite{J2021_IR2}, \citet{J1511_Radio16}, \citet{OQ208_RADIO8}, \citet{J2021_RADIO8}, \cite{J2021_IR6}, \citet{J1511_Radio17}, \citet{J1511_Radio9}, \citet{OQ208_RADIO20} \cite{OQ208_RADIO10}, \citet{OQ208_RADIO12}, \citet{OQ208_RADIO13}, \citet{J1511_Radio6} \cite{J2021_RADIO17}, \citet{J2021_RADIO18}, \citet{J2021_RADIO19}, \citet{J2021_RADIO20} \cite{J2021_RADIO21}, \citet{OQ208_RADIO25}, \citet{J2021_RADIO23}, \citet{Stanghellini98} \cite{J2021_RADIO25}, \citet{OQ208_RADIO26}, \citet{J1511_Radio8}, \citet{J1511_Radio7} \cite{OQ208_RADIO34}, \citet{OQ208_RADIO35}, \citet{OQ208_RADIO37}, \citet{J2021_RADIO32} \cite{OQ208_RADIO36}, \citet{J2021_IR5}, \citet{J2021_RADIO35} \\
  & & \\
{ Infrared}  & {J1407+2827} & \cite{J1511_IR1}, \citet{J1511_IR2}, \citet{OQ208_IR3} \\ 
& J1511+0518 & \cite{J1511_IR1}, 
\citet{J1511_IR2} \\
 & {J2022+6137} & \cite{J1511_IR2}, \citet{J2021_IR2}, \citet{J2021_IR3}, \citet{J2021_IR4}, \cite{J2021_IR5}, \citet{J2021_IR6}, \citet{J2021_IR7} \\
 & & \\
{ Optical/UV}  & {J1407+2827} & \cite{OQ208_VIS1}, \citet{OQ208_VIS2}, \citet{OQ208_VIS3}, \citet{OQ208_VIS4}, 
 \cite{J1511_VIS3}, \citet{OQ208_VIS6}, \citet{OQ208_VIS7}, \citet{J1511_UV1} \cite{J1511_UV2} \\
 & J1511+0518 & \cite{J1511_VIS1}, \citet{J1511_VIS2}, \citet{J1511_VIS3}, \citet{J1511_UV1}, \cite{J1511_UV2} \\
 & {J2022+6137} & \cite{OQ208_VIS4}, \citet{J1511_VIS3} \\\hline
\enddata
\end{deluxetable*}

\section{Basic Energetic Considerations}

\label{sec:appendix}
The accretion disk UV luminosity can be written as
\begin{equation}
    L_{\rm UV} \simeq \eta_{\rm d} \, \dot{M}_{\rm acc} c^2 \, ,
\end{equation}
where $\eta_{\rm d}$ is the disk radiative efficiency, and $\dot{M}_{\rm acc}$ is the mass accretion rate. For standard geometrically-thin/optically-thick disks, one expects $\eta_{\rm d} \sim 0.1$ \citep[depending however on the black hole spin; see, e.g.,][and references therein]{Abramowicz13}. Some fraction of this emission may be reprocessed within the disk corona, and re-emitted in the X-ray range, so that the resulting corona luminosity is $L_{\rm X} \simeq \eta_{\rm c} \, L_{\rm UV}$. The efficiency factor here, $\eta_{\rm c}$, may obviously change from source to source, and may in particular scale with the disk luminosity. A relatively tight X-ray/UV luminosity correlation established for quasars \citep[e.g.,][]{Lusso16}, indicates nevertheless average values of $\eta_{\rm c} \sim 0.1$ for lower-luminosity sources ($L_{\rm UV} \lesssim 10^{45}$\,erg\,s$^{-1}$), decreasing down to $\sim 0.01$ for high-luminosity ones ($L_{\rm UV} \gtrsim 10^{47}$\,erg\,s$^{-1}$). The mid-IR luminosity of a dusty torus is also expected to constitute some fraction of a time-averaged disk emission, $L_{\rm IR} \simeq \eta_{\rm t} \, L_{\rm UV}$, where $\eta_{\rm t} \gtrsim 0.5$ as implied by the analysis of the broad-band SEDs of quasars \citep[see, e.g.,][]{Ralowski23}. 

The energy density of the nuclear emission components, including both the torus IR and the disk UV emission, at the distances larger than the dust sublimation radius, i.e., effectively on the scales of the order of the CSO linear size LS\,$ > 1$\,pc (defined in this paper as half of the hotspot–hotspot distance), is
\begin{equation}
    U_0 = \frac{(1+ \eta_{\rm t}) \, L_{\rm UV}}{4\pi \, {\rm LS}^2 \, c} \, .
\end{equation}
The total IC luminosity of the lobes is therefore
\begin{equation}
    L_{\rm IC} = \frac{4 c \sigma_T}{3 m_e c^2} \,\, f_{\gamma} \,\, V \, U_0 \,  U_e \, ,
\end{equation}
where $V$ is the total volume of the lobes, 
\begin{equation}
f_{\gamma} \equiv \langle \gamma^2\rangle / \langle \gamma\rangle =  \int\!d\gamma \, \gamma^2 \, N_e\!(\gamma) \,\, / \, \int\!d\gamma \, \gamma \, N_e\!(\gamma) \, ,
\end{equation}
$N_e\!(\gamma)$ is the evolved electron energy distribution within the lobes, and $U_e$ is the corresponding electron energy density,
\begin{equation}
    U_e \equiv \int\!d\gamma \, \gamma m_e c^2 \, N_e\!(\gamma).
\end{equation}
For various electron injection spectra and other CSO parameters explored in LS08, $f_{\gamma}$ turned out typically of the order of the electron cooling energy, $\gamma_{\rm cr} \sim 10^2-10^3$. 

In spite of the electron energy distribution $N_e\!(\gamma)$ evolving with time in the expanding lobes due to the adiabatic and radiative cooling, LS08 argued that at every moment of the CSO lifetime, $U_e$ constitutes approximately the same fixed fraction of the lobes' total pressure $p$, namely $U_e \simeq \eta_e \, p$. In the case radiating electrons dominate the lobes' pressure, one has for example $\eta_e \simeq 3$. And since the total energy of the plasma filling the lobes is deposited by a pair of jets over the source lifetime, one may further write
\begin{equation}
3 p V = 2 L_{\rm j} \, \tau_{\rm j} \, ,
\end{equation}
where $L_{\rm j}$ is the jet kinetic power, and $\tau_{\rm j} = {\rm LS} / v_{\rm h}$ is the source lifetime for the observed (and constant, by assumption) hotspots advance velocity, $v_{\rm h}$ (defined here as half of the hotspot–hotspot separation velocity). Note that the jet total power and the mass accretion rate can be related by introducing the jet production efficiency parameter $\eta_{\rm j}$, namely
\begin{equation}
    2 L_{\rm j} = \eta_{\rm j}  \, \dot{M}_{\rm acc} c^2 \, .
\end{equation}
In principle, in the most favourable conditions provided by the magnetically arrested disks around maximally spinning black holes, $\eta_{\rm j}$ may even exceed unity, as the jet power is in this case extracted at the expense of the accretion power \emph{and} the black hole rotational energy \citep{Tchekhovskoy11}. However, for the sample of known CSOs with measured parameters of the central engine, \citet{Wojtowicz20} observed in general values of the jet production efficiency parameter much lower than unity, with a significant spread around from $\eta_{\rm j} \gtrsim 0.01$ to $\eta_{\rm j} < 0.1$.
%from $\eta_{\rm j} \gtrsim 0.1$ to $\eta_{\rm j} < 0.01$.

Taking into account the above considerations, the ratio of the lobes' total IC luminosity and the disk corona luminosity, reads as
\begin{equation}
\label{eq:xratio}
    \frac{L_{\rm IC}}{L_{\rm X} } \simeq \frac{4 \, G m_p}{9 \, m_e c} \, \frac{(1+\eta_{\rm t}) \, \eta_e \, \eta_{\rm j}}{\eta_{\rm c} \, \eta_{\rm d}} \, \frac{f_{\gamma} \, \lambda_{\rm Edd} \, M_{\rm BH}}{{\rm LS} \, v_{\rm h}} \,\, \sim \,\, \frac{(1+\eta_{\rm t}) \, \eta_e \, \eta_{\rm j}}{\eta_{\rm c} \, \eta_{\rm d}} \, \left(\frac{f_{\gamma}}{10^3} \right) \, \left(\frac{M_{\rm BH}}{10^9 M_{\odot}} \right) \, \left(\frac{\lambda_{\rm Edd}}{0.01} \right) \, \left(\frac{{\rm LS}}{10\,{\rm pc}} \right)^{-1} \, \left(\frac{v_{\rm h}}{0.03 c} \right)^{-1}     \, ,
\end{equation}
where we introduced the Eddington \text{ratio}
\begin{equation}
\lambda_{\rm Edd} \equiv \frac{L_{\rm UV}}{L_{\rm Edd}} \, ,
\end{equation}
for the Eddington luminosity $L_{\rm Edd} = 4 \pi G M_{\rm BH} m_p c / \sigma_T$ corresponding to the black hole mass $M_{\rm BH}$. The luminosity ratio given in Equation~\ref{eq:xratio} may be larger or smaller than unity, depending on a particular combination of the source parameters. In general, however, one may expect that the lobes' IC emission dominates over the disk corona radiative output in the case of the youngest and the most powerful CSOs, i.e. sources characterized by high accretion rates $\lambda_{\rm Edd} > 0.01$ and compact lobes with linear sizes LS$\,\lesssim 10$\,pc, preferentially located in a dense environment so that the jets' advance velocities are low, $v_{\rm h} < 0.1 c$.

\end{document}